\documentclass[a4paper,USenglish]{lipics-v2019}
\pdfoutput=1

\usepackage{latexsym,amsmath,amsfonts,amssymb,stmaryrd,mathtools}
\usepackage{amsthm}
\usepackage[nodate]{datetime}
\usepackage{listings}
\usepackage{graphicx}
\usepackage{verbatim}
\usepackage[T1]{fontenc}
\usepackage[normalem]{ulem}
\usepackage{booktabs}
\usepackage{float}
\usepackage[subtle]{savetrees}
\usepackage{dashrule}
\usepackage[utf8]{inputenc}
\usepackage{tikz}
\usepackage{pgfplots}
\usepackage{tablefootnote}

\newfloat{lstfloat}{!htb}{lop}
\floatname{lstfloat}{Listing}

\graphicspath{{./fig/}}

\newcommand{\cas}{\textsc{CAS}}
\newcommand{\mcas}{\textsc{MCAS}}
\newcommand{\llsc}{\textsc{LL/SC}}
\newcommand{\pmcas}{\textsc{PMCAS}}
\newcommand{\ttactive}{\texttt{ACTIVE}}
\newcommand{\successful}{\texttt{SUCCESSFUL}}
\newcommand{\failed}{\texttt{FAILED}}
\newcommand{\worddescriptor}{\texttt{WordDescriptor}}
\newcommand{\mcasdescriptor}{\texttt{MCASDescriptor}}
\newcommand{\readInternal}{\texttt{readInternal}}
\newcommand{\isDescriptor}{\texttt{isDescriptor}}
\def\ttread{\texttt{read}}

\newcommand{\ignore}[1]{}
\newcommand{\xparagraph}[1]{\smallskip\noindent\textsf{\textbf{#1}}\quad}

\newcommand{\remove}[1] {}
\newcommand{\extabstract}[1] {}

\usetikzlibrary{
    pgfplots.colorbrewer,
} 
\pgfplotsset{compat=newest}
\pgfplotsset{
    tick label style={font=\small},
    label style={font=\small},
    legend style={font=\footnotesize},
    every axis/.append style={line width=1pt},
    cycle list name=exotic,
    cycle list/.define={my marks}{
        every mark/.append style={solid,fill=\pgfkeysvalueof{/pgfplots/mark list fill}},mark=*\\
        every mark/.append style={solid,fill=\pgfkeysvalueof{/pgfplots/mark list fill}},mark=square*\\
        every mark/.append style={solid,fill=\pgfkeysvalueof{/pgfplots/mark list fill}},mark=triangle*\\
        every mark/.append style={solid,fill=\pgfkeysvalueof{/pgfplots/mark list fill}},mark=diamond*\\
    },
    mark list fill={.!75!white},
    cycle multiindex* list={
        exotic
            \nextlist
        my marks
            \nextlist
        linestyles*
            \nextlist
        very thick
            \nextlist
    },
}
\usepgfplotslibrary{groupplots}

\usepackage{versions}

\includeversion{arxiv}
\excludeversion{camera}

\title{Efficient Multi-word Compare and Swap\footnote{A shorter version of this paper will appear in the proceedings of DISC '20.}} 
\hideLIPIcs

\author{Rachid Guerraoui}{EPFL, Switzerland}{rachid.guerraoui@epfl.ch}{}{}

\author{Alex Kogan}{Oracle Labs, USA}{alex.kogan@oracle.com}{}{}

\author{Virendra J. Marathe}{Oracle Labs, USA}{virendra.marathe@oracle.com}{}{}

\author{Igor Zablotchi\footnote{This work was done when the author was an intern at Oracle Labs.}}{EPFL, Switzerland}{igor.zablotchi@epfl.ch}{}{}

\authorrunning{R. Guerraoui, A. Kogan, V.\,J. Marathe, and I. Zablotchi} 

\Copyright{Rachid Guerraoui, Alex Kogan, Virendra J. Marathe, and Igor Zablotchi} 



\acknowledgements{This work has been supported in part
by the European Research Council (ERC) Grant 339539 (AOC).}

\nolinenumbers 

\begin{camera}
\end{camera}
\begin{arxiv}
\hideLIPIcs  
\end{arxiv}

\EventEditors{Hagit Attiya}
\EventNoEds{1}
\EventLongTitle{34rd International Symposium on Distributed Computing (DISC 2020)}
\EventShortTitle{DISC 2020}
\EventAcronym{DISC}
\EventYear{2020}
\EventDate{October 12--18, 2020}
\EventLocation{Virtual Conference}
\EventLogo{}
\SeriesVolume{179}
\ArticleNo{38}

\begin{document}

\maketitle

\begin{abstract}
Atomic lock-free multi-word compare-and-swap (\mcas) is a powerful tool for designing concurrent algorithms. 
Yet, its widespread usage has been limited because lock-free implementations of \mcas{} make heavy use of expensive compare-and-swap (\cas) instructions. Existing \mcas{} implementations indeed use at least $2k+1$ \cas{}es per $k$-\cas{}. This leads to the natural desire to minimize the number of \cas{}es required to implement \mcas{}.

We first prove in this paper that it is impossible to ``pack'' the information required to perform a $k$-word \cas{} ($k$-\cas{}) in less than $k$ locations to be \cas{}ed. Then we present the first algorithm that requires $k+1$ \cas{}es per call to $k$-\cas{} in the common uncontended case. 
We implement our algorithm and show that it outperforms a state-of-the-art baseline 
in a variety of benchmarks in most considered workloads.
We also present a durably linearizable (persistent memory friendly) version of our \mcas{} algorithm using only $2$ persistence fences per call, while still only requiring $k+1$ \cas{}es per $k$-\cas{}.

\end{abstract}


\section{Introduction}
\label{sec:intro}

Compare-and-swap (\cas) is a foundational primitive used 
pervasively in concurrent algorithms on shared
memory systems. In particular, it is used extensively in \textit{lock-free} algorithms, which 
avoid the pitfalls of blocking synchronization (e.g., that employs locks) 
and typically deliver more scalable performance on multicore systems.
\cas\ conditionally updates a memory word such that a new value is written if 
and only if the old value in that word matches some expected value.
\cas\ has been shown to be universal, and thus can implement any shared object in a 
non-blocking manner~\cite{herlihy93}. This primitive (or the similar load-linked/store-conditional (\llsc)) is nowadays
provided by nearly every modern architecture.

\cas\ does have an inherent limitation: it operates on a single word.
However, many concurrent algorithms require atomic modification of
multiple words, thus introducing significant complexity (and
overheads) to get around the 1-word restriction of
\cas~\cite{braginsky12,detlefs00,greenwald99,greenwald02,levandoski13,natarajan14}.
As a way to address the 1-word limitation, the research community suggested a
natural extension of \cas\ to multiple words---an
atomic multi-word compare-and-swap (\mcas).  \mcas\ has been
extensively investigated over the last two
decades~\cite{anderson95,anderson97,detlefs00,greenwald99,greenwald02,harris02,herlihy93,moir97,shavit95}.
Arguably, this work partly led to the advent of the enormous wave of
Transactional Memory (TM) research~\cite{harris10,harris03,herlihy93a}.  In
fact, MCAS can be considered a special case of TM.
While \mcas\ is not a silver bullet for concurrent
programming~\cite{doherty04,herlihy03}, the extensive body of literature
demonstrates that the task of designing concurrent algorithms becomes
much easier with \mcas.  
Not surprisingly, there has been a resurgence
of interest in \mcas\ in the context of persistent memory, where the persistent
variant of \mcas\ (\pmcas) serves as a building block for highly
concurrent data structures, such as skip lists and
B+-trees~\cite{arulraj18,wang18}, managed in persistent memory.

Existing lock-free \mcas{} constructions typically make heavy use of \cas{} instructions~\cite{anderson95,harris02,moir97}, requiring between $2$ and $4$ \cas{}es per word modified by \mcas{}. That resulting cost is high: \cas{}es may cost up to $3.2\times$ times more cycles than load or store instructions~\cite{DavidGT13}. Naturally, algorithm designers aim to minimize the number of \cas{}es in their \mcas{} implementations. 

Toward this goal, it may be tempting to try to ``pack'' the information needed to perform the \mcas{} in fewer than $k$ memory words and perform \cas{} only on those words. 
We show in this paper that this is impossible. While this result might not be surprising, the proof is not trivial, and is done in two steps. First, we show through a bivalency argument that lock-free \mcas{} calls with non-disjoint sets of arguments must perform \cas{} on non-disjoint sets of memory locations, or violate linearizability. Building on this first result, we then show that any lock-free, disjoint-access-parallel $k$-word \mcas{} implementation admits an execution in which some call to \mcas{} must perform \cas{} on at least $k$ different locations. (Our impossibility result focuses on \textit{disjoint-access-parallel} (DAP) algorithms, in which \mcas{} operations on disjoint sets of words do no interfere with each other. DAP is a desirable property of scalable concurrent algorithms~\cite{israeli94}.) 

We also show, however, in the paper that \mcas{} can be ``efficient''. 
We present the first \mcas{} algorithm that requires $k+1$ \cas{} instructions per call to $k$-\cas{} (in the common uncontended case). Furthermore, our construction has the desirable property that reads do not perform any writes to shared memory (unless they encounter an ongoing \mcas{} operation). This is to be contrasted with existing \mcas{} constructions (in which read operations do not write) that use at least $3k+1$ \cas{}es per $k$-\cas{}. 
Furthermore, we extend our \mcas{} construction to work with persistent memory (PM). The extension does not change the number of \cas{}es and requires only 2 persistence fences per call (in the common uncontended case), comparing favorably to the prior work that employs $5k+1$ \cas{}es and $2k+1$ fences~\cite{wang18}.

Most previous \mcas{} constructions follow a multi-phase approach to perform a $k$-\cas{} operation $op$.
In the first (\textit{locking}) phase, $op$ ``locks'' its designated memory locations one by one by replacing the current value in those locations with a pointer to a \textit{descriptor} object. This descriptor contains all the information necessary to complete $op$ by the invoking thread or (potentially) by a helper thread.  In the second (\textit{status-change}) phase, $op$ changes a status flag in the descriptor to indicate successful (or unsuccessful) completion.
In the third (\textit{unlocking}) phase, $op$ ``unlocks'' those 
designated memory locations, replacing pointers to its descriptor with new or old values, depending on whether $op$ has succeeded or failed.


In order to obtain lower complexity, our algorithm makes two crucial observations concerning this unlocking phase. First, this phase can be deferred off the critical path with no impact on correctness. In our algorithm, once an \mcas{} operation completes, its descriptor is left in place until a later time. The unlocking is performed later, either 
by another \mcas{} operation locking the same memory location (and thus effectively eliminating the cost of unlocking for $op$) or during the memory reclamation of operation descriptors. (We describe a delayed memory reclamation scheme that employs epochs and amortizes the cost of reclamation across multiple operations.)


Our second, and perhaps more surprising, observation is that deferring the unlocking phase allows the \textit{locking} phase to be implemented more efficiently. In order to avoid the ABA problem, many existing algorithms require extra complexity in the locking phase. For instance, the well-known Harris et al.~\cite{harris02} algorithm uses the atomic \textit{restricted double-compare single-swap} (\textit{RDCSS}) primitive (that requires at least $2$ \cas{}es per call) to conditionally lock a word, provided that the current operation was not completed by a helping thread. Naively performing the locking phase using \cas{} instead of RDCSS would make the Harris et al.\ algorithm prone to the ABA problem (we provide an example in the full version of our paper~\cite{longversion}). However, in our algorithm, we get ABA prevention ``for free'' by using a memory reclamation mechanism to perform the unlocking phase, because such mechanisms already need to protect against ABA in order to reclaim memory safely. 

Deferring the unlocking phase allows us to come up with an elegant and, arguably, simple \mcas{} construction.  
Prior work shows, however, that the correctness of an \mcas{} construction should not be taken for granted:
for instance, Feldman et al.~\cite{feldman15} and Cepeda et al.~\cite{CepedaCLLWG19} describe correctness pitfalls in \mcas{} implementations.
In this paper, we carefully prove the correctness of our construction. 
We also evaluate our construction empirically by comparing to a state-of-the-art \mcas{} implementation and showing superior performance through a variety of benchmarks (including a production quality B+-Tree~\cite{arulraj18}) in most considered scenarios.

We note that the delayed unlocking/cleanup introduces a trade-off between higher \mcas{} performance (due to fewer \cas{}es per \mcas{}, which also leads to less slow-down due to less helping) and lower read performance (because of the extra level of indirection reads have to traverse when encountering a descriptor left in place after a completed \mcas{}). One may argue that it also increases the amount of memory consumed by the \mcas{} algorithm. Regarding the former, our evaluation shows that the benefits of the lower complexity overcome the drawbacks of indirection in all workloads that experience \mcas{} contention. Furthermore, we propose a simple optimization to mitigate the impact
of indirection in reads. As for the latter, we note that much like any lock-free algorithm, the memory consumption of our construction can be tuned by performing memory reclamation more (or less) often. 

The rest of the paper is organized as follows. In Section~\ref{sec:model} we describe our model. In Section~\ref{sec:lowerbound} we present our impossibility result.
Sections~\ref{sec:algos} and \ref{sec:persistent-mcas} detail our MCAS algorithms for volatile and persistent memory.
Section~\ref{sec:mem-mgmt} elaborates our lazy memory reclamation scheme. Section~\ref{sec:eval} presents the results of our experimental evaluation. We review related work in Section~\ref{sec:related-work} and conclude in Section~\ref{sec:conclusion}. 
\begin{camera}
Due to space limitations, some content (proofs, additional performance results etc.) has been omitted and appears in the full version of this paper~\cite{longversion}.
\end{camera}\begin{arxiv}
Some content has been moved to the optional appendices: Appendix~\ref{sec:aba-harris} provides the ABA example for the naive simplification of the Harris et al.\ algorithm; Appendix~\ref{sec:appendix-correctness} proves the correctness of our volatile algorithm; Appendix~\ref{app:persistent-mcas} presents the persistent version of our algorithm; Appendices~\ref{sec:mem-mgt-app} through Appendix~\ref{sec:reads} provide additional details regarding our memory management scheme for volatile and persistent memory and regarding an optimization to improve read performance; Appendix~\ref{sec:extra-evaluation} contains additional performance graphs and Appendix~\ref{sec:related-non-blocking-mcas} gives additional details regarding related work on non-blocking \mcas{}.
\end{arxiv}


\section{System Model}
\label{sec:model}
\subsection{Volatile Memory}
\label{sec:volatile-model}
We assume a standard model of asynchronous shared memory~\cite{HW+91}, with basic atomic \emph{read}, \emph{write} and \emph{compare-and-swap} (\cas{}) operations.
The latter receives three arguments---an address, an expected value and a new value; it reads the value stored in
the given address and if it is equal to the expected value, atomically stores the new value in the given address, returning the indication of success or failure.

\extabstract{
A program is executed by $n$ deterministic threads, where $n$ can 
be larger than the number of available processors.
A scheduler decides which threads take execution steps, when and on which processor.
We make no assumptions on the relative speeds of the processes.
In particular, at any point in time, any thread can be suspended for arbitrarily long or even 
indefinitely (e.g., if it crashes) between any of its two execution steps.

Threads communicate through atomic operations on predefined set of shared memory locations or words.
The available atomic operations are \emph{read}, \emph{write} and \emph{compare-and-swap} (\cas{}).
The latter receives three arguments---an address, an expected value and a new value; it reads the value stored in
the given address and if it is equal to the expected value, atomically stores the new value in the given address.
In that case, we say that the \cas{} operation \emph{succeeds} (or \emph{is successful}).
Otherwise, if the read value is different from the expected one, the \cas{} operation has no effect on the shared memory.
In that case, we say that the \cas{} operation \emph{fails}.
Upon its invocation, the \cas{} operation returns the indication of its success or failure.
}

Using those atomic operations, we implement an atomic \mcas{} operation with the following semantics.
The \mcas{} operation receives an array of tuples, where each tuple contains an address, an expected value and a new value.
For ease of presentation, we assume the size of the array is a known constant $N$.
(In practice, the size of the array can be dynamic, and different for every \mcas{} operation.)
The \mcas{} operation reads values stored in the given addresses, and if they all are equal to respective expected values, atomically
writes new values to the corresponding address and returns an indication of success.
Otherwise, if at least one read value is different from an expected one, the \mcas{} operation returns an indication of failure.
We also provide a custom implementation of a read operation from a memory location that can be a 
target of an \mcas{} operation (which, in the most general case, can be any shared memory location).

Our \mcas{} implementation is \emph{linearizable}~\cite{HW+91}. 
This means, informally, that each (read or \mcas{}) operation appears to take effect instantaneously 
at some point in time in the interval during which the operation executes.
In terms of progress, our \mcas{} implementation is \emph{non-blocking}.
That is, a lack of progress of any thread (e.g., due to the suspension or failure of that thread) does not 
prevent other threads from applying their operations. Furthermore, the \mcas{} implementation guarantees \emph{lock-freedom}.
That is, given a set of threads applying operations, 
it guarantees that, eventually, at least one of those threads will complete its operation.

Similar to many non-blocking algorithms, our design makes use of operation descriptors, which store information on 
existing \mcas{} operations, including the status of the operation and the array of tuples with addresses and values.
We assume each word in the shared memory can contain either a regular value or a pointer to such a descriptor.
A similar assumption has been made in prior work on \mcas{}~\cite{feldman15, harris02, sundell11, wang18}.
In practice, a single (e.g., least significant) bit can be used to distinguish between the two.

Initialization of the descriptor is done before invocation of the \mcas{} operation.
We assume that all the addresses in the descriptor are sorted in a monotonic total order.
This assumption is crucial for the liveness property of our algorithm, buy can be easily lifted by explicitly sorting the array of tuples by corresponding addresses before an 
\mcas{} operation is executed.

\subsection{Persistent Memory}
We extend the model in Section~\ref{sec:volatile-model} with standard assumptions about PM~\cite{cohen18, david18,FriedmanHMP18,izraelevitz16}.
We assume the system is equipped with
persistent shared memory that can be accessed through the same set of atomic primitives 
(read, write and \cas{}). 
The system may also be equipped with DRAM to be used as transient storage.
As in previous work~\cite{izraelevitz16}, we assume that the overall system can crash at any time
and possibly recover later. 
On such a full-system crash, we assume that the contents of persistent memory---but not those of processor caches, registers or volatile memory---are preserved. 
Moreover, threads that are active at the time of the crash are assumed to be lost
forever and replaced by new threads in case of recovery. 
After a full-system crash but before the system recovers and resumes normal execution, 
we assume a \emph{recovery} routine may be executed, in order to bring persistent 
memory-resident objects to a consistent state.
The recovery routine can be executed in a single thread, and thus it does not have to 
be thread-safe. 
Another full-system crash, however, may occur during the recovery routine.

As is standard practice~\cite{cohen18, david18, wang18}, we assume that a-priori there is no guarantee on when and 
in what order cache lines are written back to persistent memory. 
We assume the existence of two primitives to enforce such write backs. 
The first primitive is \texttt{PERSISTENT\_FLUSH(addr)}, which takes as argument 
a memory location and asynchronously writes the contents of that location to persistent memory. 
Multiple invocations of this primitive are not ordered with respect to each other and thus 
several flushes can proceed in parallel. 
Concrete examples of this primitive are \texttt{clflushopt} and \texttt{clwb}~\cite{intel}.
The second primitive is \texttt{PERSISTENT\_FENCE()}, which stalls the CPU until 
any pending flushes are committed to persistent memory.
A concrete example of this primitive is \texttt{sfence}~\cite{intel}. LOCK-prefixed instructions such as CAS also act as persistent fences~\cite{intel}.
Since persistent flushes do not stall the CPU, whereas persistent fences do, the cost of writing to persistent memory is dominated by the latter instructions and we consider the cost of the former to be negligible. 

Regarding initialization, we assume descriptor contents are made persistent before invocation of \mcas{}.

The safety criterion we use when working with persistent memory is durable linearizability~\cite{izraelevitz16}. 
Informally, an implementation of an object is durably linearizable if it is linearizable and has the following additional properties in case of a full-system crash and recovery: (1) all operations that completed before the crash are reflected in the post-recovery state and (2) if some operation $op$ that was ongoing at the time of the crash is reflected in the post-recovery state, then so are all the operations on which $op$ depends (i.e., operations whose effects $op$ observed and thus need to be linearized before $op$).

\section{Impossibility}\label{sec:lowerbound}

In this section we show that any lock-free disjoint-access-parallel (DAP) implementation of MCAS requires at least one \cas{} per modified word. Consider a call to $k$-\cas$(addr_1, \dots, addr_k$, [old and new values]). We call ${addr_1, \dots, addr_k}$ the \textit{set of targets} of the call. We also define the \textit{range} of the call in an execution $E$ to be the set of locations on which \cas{} (single-word \cas) is performed, successfully or not, during the call in $E$. 
Intuitively, we say that an \mcas{} implementation is \textit{DAP} if non-conflicting calls to $k$-\cas{} do not access the same memory locations; for the formal definition, see~\cite{israeli94}.

\begin{definition}[Star Configuration]
We say that a set $\lbrace c_0, \dots ,c_\ell \rbrace$ of calls to $k$-\cas{} are in a \emph{star configuration} if (1) the sets of targets of $c_0$ and $c_i$ are non-disjoint for all $i \in \lbrace 1,\dots,\ell\rbrace$, and (2) the sets of targets of $c_i$ and $c_j$ are disjoint for all $i\ne j\in\lbrace1,\dots,\ell\rbrace$.
\end{definition}

An example of a star configuration for $\ell = k$ is the following set of calls $\mathcal{C} = \lbrace c_0, \dots ,c_k\rbrace$, where we omit old and new values for ease of notation and we assume that addresses $a_i^{(j)}$ are all distinct:
\begin{itemize}
    \item $c_0$: $k$-\cas{}$(a_1^{(0)}, \dots, a_k^{(0)})$
    \item $c_1$: $k$-\cas{}$(a_1^{(0)}, a_2^{(1)}, \dots, a_k^{(1)})$. Call $c_1$'s set of targets intersects that of $c_0$ in $a_1^{(0)}$.
    \item $c_i$, $1\leq i \leq k$: $k$-\cas{}$(a_1^{(i)}, \dots, a_i^{(0)}, \dots, a_k^{(i)})$. Call $c_i$'s set of targets intersects that of $c_0$ in $a_i^{(0)}$ and is disjoint from the set of targets of $c_j$ for all $j\ne i, j\ne0$.
\end{itemize}

In this section, we assume without loss of generality that all calls in $\mathcal{C}$ have the correct old values for their target addresses and that each new value is distinct from its respective old value. Under these assumptions, in every execution it must be that either $c_0$ succeeds and all $c_1,\dots,c_k$ fail, or that $c_0$
 fails and all $c_1,\dots,c_k$ succeed.

We say that a state $S$ of an implementation $\mathcal{A}$ is $c_0$-valent with respect to (\textit{wrt}) some subset $C \subseteq \mathcal{C}$ if, for any call $c_i\in C$, in any execution starting from $S$ in which only $c_0$ and $c_i$ take steps, $c_0$ succeeds. Similarly, we say that a state $S$ is $C$-valent \textit{wrt} $c_0$ if, for any call $c_i\in C$, in any execution starting from $S$ in which only $c_0$ and $c_i$ take steps, $c_0$ fails. We say that a state is univalent \textit{wrt} $c_0$ and $C$ if it is $c_0$-valent or $C$-valent; otherwise it is bivalent \textit{wrt} $c_0$ and $C$. A state is critical \textit{wrt} $c_0$ and $C$ when (1) it is bivalent \textit{wrt} $c_0$ and $C$ and (2) if any process in $\lbrace c_0 \rbrace \cup C$ takes a step, the state becomes univalent \textit{wrt} $c_0$ and $C$.

Note that the initial state of $\mathcal{A}$ must be bivalent \textit{wrt} $c_0$ and any non-empty subset of $\mathcal{S}$.

\begin{lemma}\label{lemma-disjoint}
Consider a lock-free implementation $\mathcal{A}$ of $k$-\cas{} and let $\mathcal{C} = \lbrace c_0, \dots ,c_\ell \rbrace$ be a star configuration of calls to $k$-\cas{}. Then there exists an execution $E$ of $\mathcal{A}$ such that, for all $i \geq 1$, the ranges of $c_0$ and $c_i$ in $E$ are non-disjoint.
\end{lemma}

\begin{proof}
We follow a bivalency proof structure. We construct an execution in which process $p_i$ performs call $c_i$, $i\geq0$. For ease of notation, we say that ``call $c_i$ takes a step'' to mean ``process $p_i$ takes a step in its execution of $c_i$''.

The execution proceeds in stages. In the first stage, as long as some call in $\mathcal{C}$ can take a step without making the state univalent \textit{wrt} $c_0$ and any non-empty subset of $\mathcal{C}$, let that call take a step. If the execution runs forever, the implementation is not lock-free. Otherwise, the execution enters a state $S$ where no such step is possible, which must be a critical state \textit{wrt} $c_0$ and some subset $C_1 \subseteq \mathcal{C}\setminus\lbrace c_0 \rbrace$. We choose $C_1$ to be maximal, i.e., state $S$ is not critical \textit{wrt} $c_0$ and any subset of $\mathcal{C}\setminus C_1$ (otherwise, add that subset to $C_1$).

We prove in Lemma~\ref{lemma-critical} below that $c_0$ and all calls in $C_1$ are about to perform \cas{} on some common location $l_1$. We let $c_0$ perform that \cas{} step, bringing the protocol to state $S'$. By our choice of $C_1$ as maximal, $S'$ must be bivalent \textit{wrt} $c_0$ and any subset of $\mathcal{C}\setminus C_1$. The execution now enters the second stage, in which we let calls in $\mathcal{C}\setminus C_1$ take steps until they reach a critical state \textit{wrt} $c_0$ and some subset $C_2 \subseteq \mathcal{C}\setminus C_1$. By induction, we can show that eventually $c_0$ will have reached critical points \textit{wrt} all calls in $\mathcal{C}$. At the end of the execution, we resume each process in $\mathcal{C}\setminus c_0$ for one step; they were each about to perform a \cas{} step on some location on which $c_0$ has already performed a \cas{} step. Thus, in this execution, all calls in $\mathcal{C}\setminus c_0$ have performed a \cas{} on a common location with $c_0$.
\end{proof}


\begin{lemma}\label{lemma-critical}
Consider a lock-free implementation $\mathcal{A}$ of $k$-\cas{} and let $\mathcal{C} = \lbrace c_0, \dots ,c_k \rbrace$ be a star configuration of calls to $k$-\cas{}. If $S$ is a critical state of $\mathcal{A}$ \textit{wrt} $c_0$ and some subset $C\subseteq\mathcal{C}$, then in $S$, $c_0$ and all calls in $C$ are about to perform a \cas{} step on a common location $l$.
\end{lemma}
\begin{proof}
From $S$, we consider the next steps of $c_0$ and any $c_i \in C$:

\begin{description}
\item[Case 1] One of the calls is about to read; assume \textit{wlog} it is $c_0$. Consider two possible scenarios. First scenario: $c_i$ moves first and runs solo until it returns ($c_i$ must succeed because $c_i$ took the first step). Second scenario: $c_0$ moves first and reads, then $c_i$ runs solo until it returns ($c_i$ must fail because $c_0$ took the first step). But the two scenarios are indistinguishable to $c_i$, thus $c_i$ must either succeed in both or fail in both, a contradiction.
\item[Case 2] Both calls are about to write. In this case, they must be about to write to the same register $r$, otherwise their writes commute. First scenario: $c_0$ writes $r$, then $c_i$ writes $r$, then $c_i$ runs solo until it returns ($c_i$ must fail since $c_0$ took the first step). Second scenario: $c_i$ writes $r$ and then runs solo until it returns ($c_i$ must succeed since $c_i$ took the first step). But the two scenarios are indistinguishable to $c_i$, since its write to $r$ obliterated any potential write by $c_0$ to $r$, so $c_i$ must either succeed in both scenarios or fail in both; a contradiction.
\item[Case 3] $c_0$ is about to \cas{} and $c_i$ is about to write (or vice-versa). In this case, their operations must be to the same memory location $r$ (otherwise they commute). First scenario: $c_0$ \cas{}es $r$, then $c_i$ writes to $r$ and then runs solo until $c_i$ returns ($c_i$ must fail since $c_0$ took the first step). Second scenario: $c_i$ writes to $r$ and then runs solo until it returns ($c_i$ must succeed since $c_i$ took the first step). But the two scenarios are indistinguishable to $c_i$, since its write to $r$ obliterated any preceding \cas{} by $c_0$ to $r$; thus $c_i$ must either succeed in both scenarios or fail in both; a contradiction.
\item[Case 4] Both calls are about to \cas{}. In this case, they must be about to \cas{} the same location, otherwise their \cas{}es commute. \qedhere
\end{description}
\end{proof}

\begin{theorem}
Consider a lock-free disjoint-access-parallel implementation $\mathcal{A}$ of $k$-\cas{} in a system with $n>k$ processes. Then there exists some execution $E$ of $\mathcal{A}$ such that in $E$ some call to $k$-\cas{} performs \cas{} on at least $k$ locations.
\end{theorem}

\begin{proof}
We prove the theorem by contradiction. We first assume that calls to $k$-\cas{} perform \cas{} on \textit{exactly} $k-1$ locations and derive a contradiction; we later show how assuming that $k$-\cas{} performs \cas{} on \textit{at most} $k-1$ locations also leads to a contradiction.

We construct an execution $E$ in which two concurrent but non-contending $k$-\cas{} calls (i.e., two $k$-\cas{} calls with disjoint sets of targets) perform \cas{} on the same location, thus contradicting the disjoint-access-parallelism (DAP) property and proving the theorem.

Let $c_0,\dots,c_k$ be $k+1$ calls to $k$-\cas{} in a star configuration. By Lemma~\ref{lemma-disjoint}, there exists an execution $E$ of $\mathcal{A}$ such that, for all $i \geq 1$, the ranges of $c_0$ and $c_i$ in $E$ are non-disjoint.

Let $l_1, \dots, l_{k-1}$ be the range of $c_0$. By Lemma~\ref{lemma-disjoint}, in $E$ the range of $c_1$ must intersect that of $c_0$ in at least one location; assume \textit{wlog} it is $l_1$. Furthermore, the range of $c_2$ must also intersect that of $c_0$ in at least one location; moreover, due to the DAP property, the intersection must contain some location other than $l_1$, since $c_1$ and $c_2$ have disjoint sets of targets. By induction, we can show that the range of each call $c_i, i\in\lbrace 1,2,\dots,k-1\rbrace$ intersects the range of $c_0$ in $l_i$. However, the range of $c_k$ must also intersect the range of $c_0$ in some location other than $l_1, \dots, l_{k-1}$, due to the DAP property. We have reached a contradiction.

If we now assume that calls to $k$-\cas{} perform \cas{} on $k-1$ \textit{or fewer} locations, then we also reach a similar contradiction as above. In fact, if some call $c_i$ performs \cas{} on strictly fewer than $k-1$ locations, this may cause the contradiction to occur before call $c_k$, as $c_i$ now has fewer locations to choose from in order intersect with the range of $c_0$ in some location that is not in the ranges of $c_1,\dots, c_{i-1}$.
\end{proof}

\section{Volatile \mcas{} with \texorpdfstring{$k+1$}{k+1} \cas{}}
\label{sec:algos}

In this section we describe our \mcas\ construction for volatile memory.
Our algorithm uses $k+1$ \cas{} operations in the common uncontended case, and does not involve cleaning up after completed \mcas\ operations.
In Section~\ref{sec:mem-mgmt} we describe a memory management scheme that can
be used to clean up after completed \mcas\ operations as well as for reclaiming or reusing
operation descriptors employed by the algorithm.

\subsection{High-level Description}

As is standard practice~\cite{ha04, harris02, sundell11}, our \mcas{} construction supports two operations: \mcas\ and \ttread. 
Similarly to most \mcas\ algorithms~\cite{ha04, harris02, sundell11}, the \mcas\ operation uses operation descriptors that contain
a set of addresses (the \emph{target} addresses or words), and \emph{old} and \emph{new} values for each target address.
In addition, each operation descriptor contains a \emph{status} word indicating the status of the corresponding \mcas\ operation.

The \mcas\ operation proceeds in two stages. 
In the first stage, we attempt to install a pointer to the operator descriptor in each memory word 
targeted by the \mcas\ operation.
If we succeed to install the pointer, we say that the target address is \emph{owned} (or \emph{locked}) by the descriptor.
The first stage ends when all target addresses are owned by the descriptor, or if we find
a target address with a value different from the expected one.
In the second stage, we \emph{finalize} the \mcas\ operation by atomically changing its status 
to indicate its success or failure, depending on whether the first stage was successful (i.e., all target addresses have been locked).
The \ttread\ operation returns the current value at an address, either by reading it directly from the target address 
or by reading the appropriate value from a descriptor of a completed \mcas\ operation installed in that address.
If either \mcas\ or \ttread\ encounter another \mcas\ in progress (e.g., when they attempt to
read the current value in the target address), they first help that \mcas\ operation to complete.


\subsection{Technical Details}

\xparagraph{Structures and Terminology.}
We describe the structures used by our algorithm and explain the terminology. 
Pseudocode for the structures is shown in Listing~\ref{list:struct}. 
An \mcasdescriptor\ describes an \mcas{} operation. It contains a status field, which can be \ttactive, \successful\ or \failed, 
the number \texttt{N} of words targeted by the \mcas{} and an array of {\worddescriptor}s for those words. 
These {\worddescriptor}s are the \emph{children} of the \mcasdescriptor, who is their \emph{parent}. 
We say that an \mcasdescriptor\ (and the \mcas\ it describes) is \emph{active} if its status is \ttactive\ and 
\emph{finalized} otherwise. 

The \worddescriptor\ contains information related to a given word as target of an \mcas{} operation: the word's address in memory, its expected value and the new intended value. 
The \worddescriptor\ also contains a pointer to the descriptor of its parent \mcas\ operation.
As described later, the pointer is used as an optimization for fast lookup of the status field in the \mcasdescriptor, and can be eliminated.

\begin{lstlisting}[caption={Data structures used by our algorithm},label=list:struct,numbers=none,float,keywords={}]
struct WordDescriptor {
    void* address;
    uintptr_t old;
    uintptr_t new;
    MCASDescriptor* parent; };
    
enum StatusType { ACTIVE, SUCCESSFUL, FAILED };

struct MCASDescriptor {
    StatusType status;
    size_t N;
    WordDescriptor words[N]; };
\end{lstlisting}

\begin{lstlisting}[caption={The \readInternal\ auxiliary function, used by our algorithm.},label=list:readInternal,float,keywords={}]
readInternal(void* addr, MCASDescriptor *self) {@\label{r_i_begin}@
retry_read:							
	val = *addr;@\label{read-internal-first-read}@
	if (!isDescriptor(val))@\label{read-internal-line4}@ then return <val,val>;@\label{read-internal-line5}@
	else { // found a descriptor@\label{read-internal-line6}@
		MCASDescriptor* parent = val->parent;@\label{read-internal-line7}@
		if (parent != self && parent->status == ACTIVE) {@\label{read-internal-check}@
			MCAS(parent);@\label{mcas-help}@
			goto retry_read;@\label{read-internal-retry}@
		} else {@\label{read-internal-line11}@
			return parent->status == SUCCESSFUL ?
			    <val,val->new> : <val,val->old>;@\label{read-internal-line12}@ }  }  }@\label{r_i_end}@
\end{lstlisting}

\begin{lstfloat}
\begin{lstlisting}[caption={Our main algorithm. Commands in {\it italic} are related to memory reclamation (discussed in a later section).},label=list:algo,firstnumber=last,keywords={}]
read(void* address) { @\label{read_begin}@
	@{\it epochStart();}@@\label{read-epoch-start}@
	<content, value> = readInternal(address, NULL);@\label{read-line2}@
	@{\it epochEnd();}@@\label{read-epoch-end}@
	return value; } @\label{read_end}@

MCAS(MCASDescriptor* desc) {
	@{\it epochStart();}@
	success = true;
	for wordDesc in desc->words {@\label{mcas-first-phase-begin}@
retry_word:
		<content, value> = readInternal(wordDesc.address, desc); @\label{mcas-readinternal}@
		// if this word already points to the right place, move on
		if (content == &wordDesc) continue;@\label{disregarded}@
		// if the expected value is different, the MCAS fails
		if (value != wordDesc.old) {@\label{doomed-start}@ success = false; break; @\label{for-break}@ }@\label{doomed-end}@
		if (desc->status != ACTIVE) break; @\label{status-check}@
		// try to install the pointer to my descriptor; if failed,	retry
		if (!CAS(wordDesc.address, content, &wordDesc)) @\label{mcas-acquire}@goto retry_word;@\label{mcas-retry}@ }@\label{mcas-first-phase-end}@
	if (CAS(&desc.status, ACTIVE, success ? SUCCESSFUL : FAILED)){@\label{mcas-finalize}@
		// if I finalized this descriptor, mark it for reclamation
		@{\it retireForCleanup(desc);}@@\label{retire-for-cleanup}@ }@\label{mcas-finalize-end}@
	returnValue = (desc.status == SUCCESSFUL);
	@{\it epochEnd();}@
	return returnValue; }
\end{lstlisting}
\end{lstfloat}

\xparagraph{Algorithm.}
Both \mcas\ and \ttread\ operations rely on the auxiliary \readInternal\ function shown in Listing~\ref{list:readInternal}. 
The \readInternal\ function takes an address \texttt{addr} and an \mcasdescriptor\ \texttt{self} (called the \emph{current descriptor}) and returns a tuple. 
The tuple contains two values (which might be identical), and, intuitively, represent the contents in the given (target) address and the actual value 
the former represents.
More specifically, \readInternal\  reads the content of the given \texttt{addr} (Line~\ref{read-internal-first-read}). 
If \texttt{addr} does not point to a descriptor (this is determined by the \isDescriptor\ function; see below), the returned tuple contains two copies of the contents of \texttt{addr} (Line~\ref{read-internal-line5}).
If \texttt{addr} points to an active \worddescriptor\ whose parent is not the same as \texttt{self}, then \readInternal\ helps the other (\mcas) 
operation to complete (Line~\ref{mcas-help}) and then restarts (Line~\ref{read-internal-retry}).
Therefore, the role of the \texttt{self} pointer is to avoid an (\mcas) operation to help itself recursively.
If \texttt{addr} points to a finalized descriptor, the tuple returned by \readInternal\ contains 
the pointer to the descriptor and the final value, corresponding to the status of the descriptor (Line~\ref{read-internal-line12}).
Finally, if \texttt{addr} points to a descriptor whose parent is equal to \texttt{self}, 
then \readInternal\ returns the pointer to that descriptor (Line~\ref{read-internal-line12}; a value is also returned in the tuple in this case, but is disregarded; see below).

Listing~\ref{list:algo} provides the pseudo-code for the \ttread\ and \mcas\ operations.
The pseudo-code includes extensions relevant to memory management (in \emph{italics}), whose discussion is deferred to Section~\ref{sec:mem-mgmt}.

The \ttread\ operation is simply a call to \readInternal\ with a \texttt{self} equal to \texttt{null} as the current operation descriptor (Line~\ref{read-line2}).

The \mcas\ operation takes as argument an \mcasdescriptor\ and returns a boolean indicating success or failure. 
As mentioned above, the operation proceeds in two stages. 
In the first stage, \mcas\ attempts to take ownership of (or \textit{acquire}) each target word (Lines~\ref{mcas-first-phase-begin}--\ref{mcas-first-phase-end}).
To this end, for each \worddescriptor\ $w$ in its \texttt{words} array, we start by calling \readInternal\ on $w$'s target address \texttt{addr} 
(Line~\ref{mcas-readinternal}; as described above, this handles any helping required in case another active operation owns \texttt{addr}). 
If \texttt{addr} is already owned by the current \mcas, we move on to the next word (Line~\ref{disregarded}). 
Otherwise, if the current value at \texttt{addr} does not match the expected value of $w$, 
the \mcas\ cannot succeed and thus we can skip the next {\worddescriptor}s and go to the second stage (Line~\ref{doomed-end}). 
If the values do match, we re-check if the operation is still active (line~\ref{status-check}); otherwise we go to the second stage---this prevents a memory location from being re-acquired by the current operation $op$ in case $op$ was already finalized by a helping thread. Finally, we attempt to take ownership of \texttt{addr} through a \cas{} (Line~\ref{mcas-acquire}).
Note that the failure of this \cas{} might mean that another thread has concurrently helped this \mcas\ to lock the target word.
Therefore, we simply retry taking ownership on this target word, rather than failing the \mcas\ operation (Line~\ref{mcas-retry}).

In the second stage (Lines \ref{mcas-finalize}--\ref{mcas-finalize-end}), \mcas\ finalizes the descriptor by atomically changing its status from \ttactive\ to \successful\ (if all word acquisitions were successful in stage one) or to \failed\ (otherwise).

Our pseudocode assumes the existence of the \isDescriptor\ function, which takes a value and returns \texttt{true} if and only if the value is a pointer to a \worddescriptor.  This function can be implemented, for instance, by designating a low-order \emph{mark bit} in a word to indicate whether it contains a pointer to a descriptor or not~\cite{harris02,wang18}. 
\begin{camera}
\end{camera}\begin{arxiv}
Whenever we make an address point to a descriptor (e.g., Line~\ref{mcas-acquire}) or convert the contents of a word into a pointer to descriptor (e.g., Line~\ref{read-internal-line7}), 
we also set or unset the mark bit, respectively. 
In the interest of clarity, we do not show the implementation of \isDescriptor\ or the code for marking/unmarking pointers.
\end{arxiv}

\begin{camera}
We give a proof of correctness for our algorithm in the full version of our paper~\cite{longversion}.
\end{camera}\begin{arxiv}
We give a proof of correctness for our algorithm in the Appendix.
\end{arxiv}

\section{Persistent \mcas{} with \texorpdfstring{$k+1$}{k+1} \cas{} and 2 Persistent Fences}\label{sec:persistent-mcas}

We discuss the modifications required to make our volatile \mcas{} algorithm work with persistent memory. 
\begin{camera}
In the full version of our paper~\cite{longversion}, we give the complete pseudocode for these modifications.
\end{camera}\begin{arxiv}
The extra instructions are shown underlined in Listings~\ref{list:algo-persistent} and \ref{list:readInternal-persistent} in the Appendix.
\end{arxiv}

In the \texttt{MCAS} function\begin{camera}\end{camera}\begin{arxiv}~(Listing~\ref{list:algo-persistent})\end{arxiv}, after all target locations have been successfully acquired, we add one persistent flush per target word and one persistent fence overall. The persistent fence ensures that all target locations persistently point to their respective {\worddescriptor}s before attempting to modify the status. 

When finalizing the status, we mark the status with a special \texttt{DirtyFlag}. This flag indicates that the status is not yet persistent. We then perform a persistent flush and fence after the status has been finalized. This ensures that the finalized status of the descriptor is persistent before returning from the \mcas{}. Finally, we unset the \texttt{DirtyFlag} with a simple store\begin{camera}\end{camera}\begin{arxiv}~(line~\ref{unset-dirty})\end{arxiv}; this store cannot create a race with the \cas{} that finalizes the status\begin{camera}\end{camera}\begin{arxiv}~(line~\ref{pmcas-finalize})\end{arxiv} because that \cas{} must fail (the status must be already finalized if some thread is already attempting to unset the dirty flag\begin{camera}\end{camera}\begin{arxiv}~(line~\ref{unset-dirty})\end{arxiv}).

We also modify the \texttt{readInternal} function\begin{camera}\end{camera}\begin{arxiv}~(Listing~\ref{list:readInternal-persistent})\end{arxiv} such that, when an operation $op$ encounters another operation $op'$ whose status is finalized but still has the \texttt{DirtyFlag} set, $op$ helps $op'$ persist its status and unsets the \texttt{DirtyFlag} on $op'$ status.

Our modifications enforce the following invariants. First, at the time when a descriptor becomes finalized, its acquisitions of target locations are persistent. Second, at the time when an \mcas{} operation returns, its finalized status is persistent. Third, when a read or \mcas{} operation $op$ returns, all operations on which $op$ depends are finalized and their statuses are persistent. With these invariants, we can argue that our persistent \mcas{} is correct. By correctness we refer to lock-freedom (liveness) and durable linearizability (safety). Lock-freedom is clearly preserved by our additions, thus we focus on durable linearizability. We examine the point in time when a full-system crash may occur during the execution of an \mcas{} operation $op$. There are two possibilities to consider:
\begin{enumerate}
    \item If the crash occurs before $op$'s status was finalized and made persistent, then we know that no operation $op'$ which observed the effects of $op$ could have returned before the crash; otherwise, $op'$ would have helped $op$ and persisted its status. In this case, neither $op$ nor any such $op'$ will be linearized before the crash; during recovery, their effects will be rolled back by reverting any acquired locations to their old values.
    \item If the crash occurs after $op$'s status was finalized and made persistent, then $op$ is linearized before the crash. During recovery, any locations still acquired by $op$ will be detached and given either their new or old values (depending on $op$'s success or failure status), as specified in $op$'s descriptor.
\end{enumerate}

In sum, the recovery procedure of our algorithm is as follows. The recovery goes through each operation descriptor $D$. If $D$'s status is not finalized, then we roll $D$ back by going through each target location $\ell$ of $D$; if $\ell$ is acquired by $D$ (i.e., points to $D$), then we write into $\ell$ its old value, as specified in $D$. If $D$'s status is finalized, then we detach $D$ and install final values; we go through each target location $\ell$ of $D$; if $\ell$ is acquired by $D$ and $D$ was successful (resp. failed), then we write into $\ell$ the new (resp. old) value as specified in $D$.

\section{Memory Management}
\label{sec:mem-mgmt}

The \mcas{} algorithm has been presented so far under the assumption that no memory is ever reclaimed.
For practical considerations, however, one should to be able to reclaim and/or reuse \mcas{} operation descriptors.
While efficient memory management of concurrent data structures remains an active area of research 
(see, e.g.,~\cite{alistarh17,Brown15, DHK16,poter18,wen18}),
here we describe one possible mechanism suitable for an \mcas{} implementation. 
\begin{camera} 
Due to space limitations, we briefly outline the mechanism here and defer its full description, as well as optimizations for persistent memory and efficient reads, to the extended version of our paper~\cite{longversion}.
\end{camera}\begin{arxiv}
We briefly outline the mechanism here and defer its full description, as well as optimizations for persistent memory and efficient reads, to Appendices~\ref{sec:mem-mgt-app} through~\ref{sec:reads}.
\end{arxiv}

We note that the life cycle of an operation descriptor comprises several phases.
Once its status is no longer \texttt{ACTIVE}, the (finalized) descriptor cannot be recycled just yet as certain memory locations can point to it.
Therefore, we need first to \emph{detach} such a descriptor by replacing the pointers to the descriptor (using \cas{}) with actual values 
(respective to whether the corresponding \mcas{} has succeeded or failed) in affected memory locations.
Only after that, a detached descriptor can be recycled, provided no concurrently running thread holds a
reference to it.
Note that \cas{}es in the detachment phase are necessary only for those affected memory locations that still point to the to-be-detached descriptor, which, as our evaluation shows, is rare in practice.

Our scheme keeps track of two categories of descriptors: (1) those that have been finalized but not yet detached and (2) those that have been detached but to which other threads might still hold references. Similar to RCU approaches~\cite{McKenney17, MS98}, we use thread-local epoch counters to track threads' progress and infer when a descriptor can be moved from category (1) to category (2), and when a descriptor from category (2) can be reclaimed.

\section{Evaluation}\label{sec:eval}

\subsection{Experimental Setup}

We evaluate our algorithm on a 2-socket Intel Xeon machine with two E5-2630 v4 processors operating at 3.1 GHz. Each processor has 10 cores, each core has 2 hardware threads (40 hardware threads total). Each experimental run lasts 5 seconds; shown values are the average of 5 runs. We base our evaluation on the framework available from the authors of PMwCAS~\cite{pmwcascode,wang18}.

The baseline of our evaluation is the volatile version of PMwCAS~\cite{pmwcascode,wang18}, a state-of-the-art implementation of the Harris et al.~\cite{harris02} algorithm. Like the Harris et al. algorithm, volatile PMwCAS requires $3k+1$ \cas{}es per $k$-\cas{}. We use PMwCAS as our baseline since (1) it has recent, openly available and well-maintained code and (2) it is to our knowledge the only other \mcas{} algorithm in which readers do not write to shared memory in the common uncontended case.

PMwCAS implements an optimization of the Harris et al. algorithm: it marks pointers with a special \textit{RDCSS} flag instead of allocating a distinct RDCSS descriptor. However, we found that this optimization made the PMwCAS algorithm incorrect, due to an ABA vulnerability. In our evaluation, we fixed the PMwCAS implementation to allocate and manually manage RDCSS descriptors.

Our evaluation uses three benchmarks: an \textit{array benchmark} in which threads perform \mcas{}-based read-modify-write operations at random locations in an array, a \textit{doubly-linked list benchmark}, in which threads perform \mcas{}-based operations on a list implementing an ordered set, and a \textit{B+tree benchmark} in which threads perform \mcas{}-based operations on a B+-tree. The first two benchmarks are based on the implementation available in~\cite{pmwcascode}, and the third is based on PiBench~\cite{pibencode} and BzTree~\cite{arulraj18,bztreecode}. We note, however, that we modified the
benchmark in~\cite{pmwcascode} so all threads operate on the same key range (rather than having each thread using a unique set of keys), so we could induce contention by controlling the size of the key range.

In each experiment, we vary the number of threads from 1 to 39 (we reserve one hardware thread for the main thread). Threads are assigned according to the default settings in the evaluation frameworks used~\cite{bztreecode,pibencode,pmwcascode}. In the array and list benchmarks, threads are assigned in the following way: we first populate the first hardware thread of each core on the first socket, then on the second socket, then we populate the second hardware thread on each core on the first socket, and finally the second hardware thread on each core on the second socket. The B+-tree benchmark uses OpenMP~\cite{openmp}, which dictates thread assignment; it also employs a scalable memory allocator~\cite{AfekDM11}.

\subsection{Array Benchmark}

The benchmark consists of each thread performing the following in a tight loop: reading $k$ locations at random from the array ($k=4$ in our experiments), computing a new value for each location, and attempting to install the new values using an \mcas{}.

In this benchmark we measure two quantities. The first is throughput: the number of read-modify-write operations completed successfully per time unit. The second metric is the \textit{helping ratio}. We measure the helping ratio by dividing the number of \textit{ongoing} \mcas{} operations encountered (and helped) during read or \mcas{} operations by the total number of \mcas{} operations. A higher helping ratio thus means more operations are slowed down due to the need to help other, incomplete \mcas{} operations.

\begin{figure*}
\centering
\begin{tikzpicture}
	
    \begin{groupplot}[
        group style=    {group name=my plots,
                        group size=3 by 2,
                        horizontal sep=0.7cm,
                        vertical sep=1.3cm},
        height=3.6cm,
        width=5.2cm,
        ytick scale label code/.code={},
        legend columns=-1,
        xlabel={Number of Threads},
    ]
]
        
    \nextgroupplot[
        ylabel = {Throughput (Mops/s)},
        legend entries={AOPT,PMWCAS},
        legend to name=CommonLegend1,
    ]
    \addplot table[x=Threads,y=AOPT] {data/array/pmwcas-array_size-10-absolute-perf.log};
    \addplot table[x=Threads,y=PMWCAS] {data/array/pmwcas-array_size-10-absolute-perf.log};
    \nextgroupplot
    \addplot table[x=Threads,y=AOPT] {data/array/pmwcas-array_size-100-absolute-perf.log};
    \addplot table[x=Threads,y=PMWCAS] {data/array/pmwcas-array_size-100-absolute-perf.log};
    \nextgroupplot
    \addplot table[x=Threads,y=AOPT] {data/array/pmwcas-array_size-1000-absolute-perf.log};
    \addplot table[x=Threads,y=PMWCAS] {data/array/pmwcas-array_size-1000-absolute-perf.log};
    
    \nextgroupplot[ylabel = {Helping Ratio}]
    \addplot table[x=Threads,y=AOPT] {data/array/pmwcas-array_size-10-helping-ops.log};
    \addplot table[x=Threads,y=PMWCAS] {data/array/pmwcas-array_size-10-helping-ops.log};
    \nextgroupplot
    \addplot table[x=Threads,y=AOPT] {data/array/pmwcas-array_size-100-helping-ops.log};
    \addplot table[x=Threads,y=PMWCAS] {data/array/pmwcas-array_size-100-helping-ops.log};
    \nextgroupplot
    \addplot table[x=Threads,y=AOPT] {data/array/pmwcas-array_size-1000-helping-ops.log};
    \addplot table[x=Threads,y=PMWCAS] {data/array/pmwcas-array_size-1000-helping-ops.log};

    \end{groupplot}
	
    \node[anchor=south,yshift=0.2cm] at ($(my plots c1r1.north)$){\large \bf Array Size 10};
    \node[anchor=south,yshift=0.2cm] at ($(my plots c2r1.north)$){\large \bf Array Size 100};
    \node[anchor=south,yshift=0.2cm] at ($(my plots c3r1.north)$){\large \bf Array Size 1000};

\end{tikzpicture}
\\

\ref{CommonLegend1}
\caption{Array benchmark. Top row shows throughput (higher is better), bottom row shows helping ratio (lower is better). Each column corresponds to a different array size (10, 100 and 1000, respectively).}
\label{fig:array}
\end{figure*}
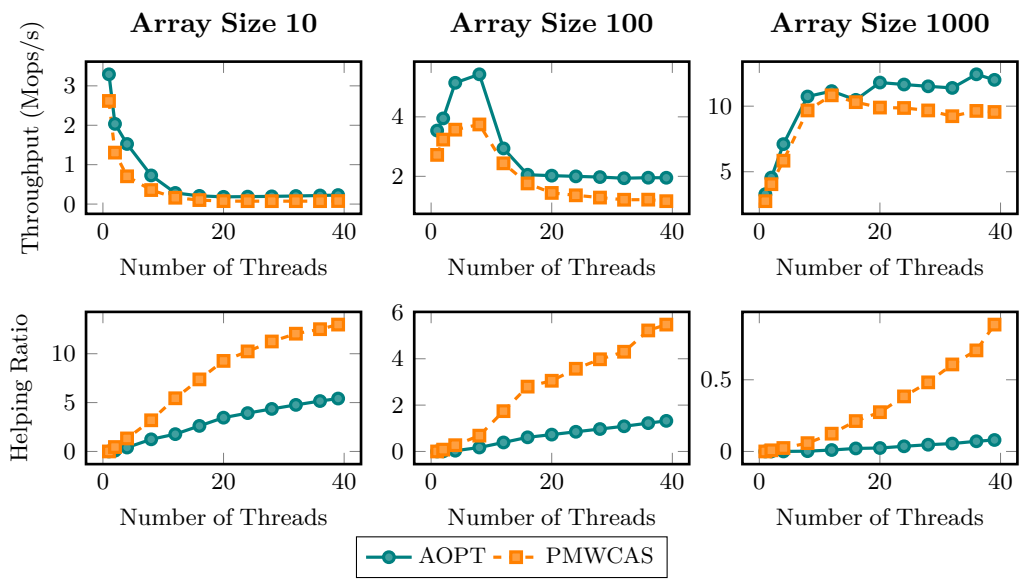

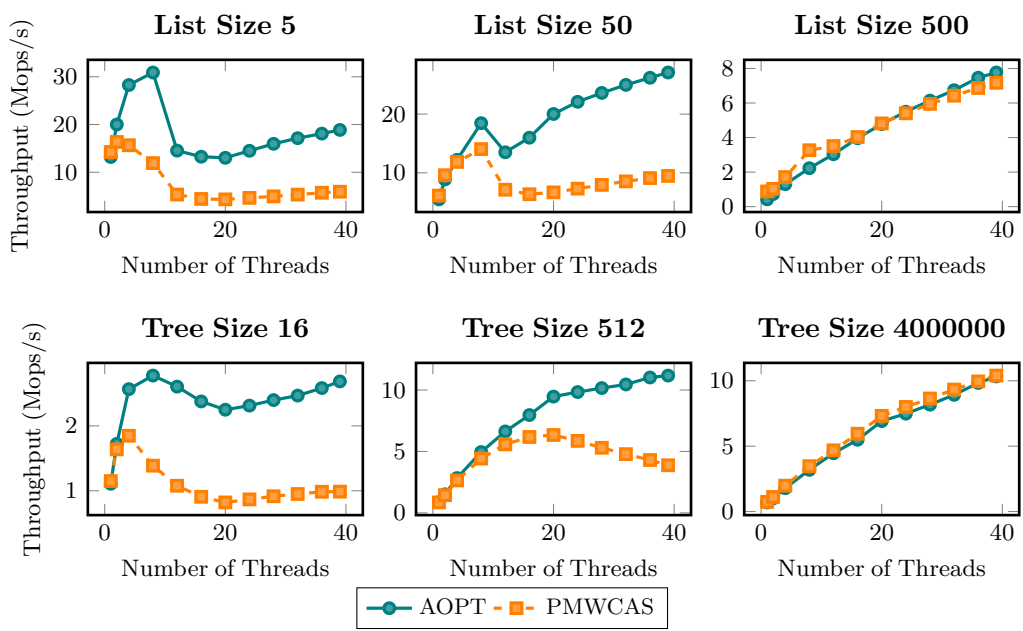
\begin{figure*}
\centering
\begin{tikzpicture}
	
    \begin{groupplot}[
        group style=    {group name=my plots,
                        group size=3 by 2,
                        horizontal sep=0.7cm,
                        vertical sep=2cm},
        height=3.6cm,
        width=5.2cm,
        ytick scale label code/.code={},
        legend columns=-1,
        xlabel={Number of Threads},
    ]
        
    \nextgroupplot[
        ylabel = {Throughput (Mops/s)},
        legend entries={AOPT,PMWCAS,AOPT-unsafe},
        legend to name=CommonLegend2,
        title = {\large \bf List Size 5},
    ]
    \addplot table[x=Threads,y=AOPT] {data/dl-list/dl-list-list_size-5-search_pct-80-absolute-perf.log};
    \addplot table[x=Threads,y=PMWCAS] {data/dl-list/dl-list-list_size-5-search_pct-80-absolute-perf.log};
    \nextgroupplot[title = {\large \bf List Size 50}]
    \addplot table[x=Threads,y=AOPT] {data/dl-list/dl-list-list_size-50-search_pct-80-absolute-perf.log};
    \addplot table[x=Threads,y=PMWCAS] {data/dl-list/dl-list-list_size-50-search_pct-80-absolute-perf.log};
    \nextgroupplot[title = {\large \bf List Size 500}]
    \addplot table[x=Threads,y=AOPT] {data/dl-list/dl-list-list_size-500-search_pct-80-absolute-perf.log};
    \addplot table[x=Threads,y=PMWCAS] {data/dl-list/dl-list-list_size-500-search_pct-80-absolute-perf.log};
    
    \nextgroupplot[ylabel = {Throughput (Mops/s)},
        title={\large \bf Tree Size 16},
    ]
    \addplot table[x=Threads,y=AOPT] {data/bztree/bztree-tree_size-16-search_pct-80-absolute-perf.log};
    \addplot table[x=Threads,y=PMWCAS] {data/bztree/bztree-tree_size-16-search_pct-80-absolute-perf.log};
    \nextgroupplot[title = {\large \bf Tree Size 512}]
    \addplot table[x=Threads,y=AOPT] {data/bztree/bztree-tree_size-512-search_pct-80-absolute-perf.log};
    \addplot table[x=Threads,y=PMWCAS] {data/bztree/bztree-tree_size-512-search_pct-80-absolute-perf.log};
    \nextgroupplot[title = {\large \bf Tree Size 4000000}]
    \addplot table[x=Threads,y=AOPT] {data/bztree/bztree-tree_size-4000000-search_pct-80-absolute-perf.log};
    \addplot table[x=Threads,y=PMWCAS] {data/bztree/bztree-tree_size-4000000-search_pct-80-absolute-perf.log};
    
    \end{groupplot}
	



\end{tikzpicture}
\\

\ref{CommonLegend2}
\caption{Top row: Doubly-linked list benchmark (80\% reads) with different initial list sizes (5, 50 and 500 elements). Bottom row: B+-tree benchmark (80\% reads) with different initial tree sizes (16, 512 and 4000000 elements).}
\label{fig:dll}
\end{figure*}

We run the benchmark with three array sizes (10, 100, and 1000) in order to capture different contention levels. The results of this benchmark are shown in Figure~\ref{fig:array} (our algorithm is denoted \textit{AOPT} in all figures in this section). 

The top row of Figure~\ref{fig:array} shows that our algorithm outperforms PMwCAS at every contention level and at every thread count, including in single-threaded mode. This can be explained by two related factors. First, our algorithm has a  lower \cas{} complexity ($k+1$ \cas{}es per $k$-\cas{} for our algorithm compared to $3k+1$ for PMwCAS). Second, as a consequence of its lower complexity, in our algorithm there is a shorter ``window'' for each \mcas{} operation to interfere with other operations by forcing them to help.

To illustrate the second factor above, we examine the helping ratios of the two algorithms (bottom row of Figure~\ref{fig:array}). We observe that the helping ratio of our algorithm is considerably lower than that of PMwCAS. This means that, on average, each operation helps (and is slowed down by) fewer \mcas{} operations in our algorithm than in PMwCAS. 

In order to quantify the impact of descriptor cleanup on performance in our algorithm, we also measure the \textit{detaching ratio}: the number of \cas{}es performed in order to detach (in the sense of Section~\ref{sec:mem-mgmt})  finalized \mcas{} descriptors, divided by the total number of completed \mcas{} operations. We find the detaching ratio to be less than $0.001$ for every thread count and array size. This is because finalized \mcas{} descriptors are constantly being replaced by ongoing \mcas{} operations, and thus recycling these detached descriptors requires no \cas{}es. We conclude that the vast majority of our \mcas{} operations do not incur any cleanup \cas{}es.

\subsection{Doubly-linked List Benchmark}

In this benchmark we operate on a shared ordered set object implemented from a doubly-linked list. The list supports search and update (insert and delete) operations. Insertions are done using 2-\cas{} and deletions are done using 3-\cas{}. We initialize the list by inserting a predefined (configurable) number of nodes. During the benchmark, each thread selects an operation type (search, insert or delete) at random, according to a configurable distribution; the thread also selects a value at random; it then performs the selected operation with the selected value.

We perform this benchmark with three initial list sizes (5, 50 and 500 elements). The operation distribution is: 80\% reads, 20\% updates (in all our experiments, updates are evenly distributed among insertions and deletions). As is standard practice, the initial size of the list is half of the key range. Results are shown in the top row of Figure~\ref{fig:dll}. We also ran experiments with 50\%, 98\%, and 100\% reads;
\begin{camera}
performance graphs for these less representative cases are available in the full version of our paper~\cite{longversion}. 
\end{camera}\begin{arxiv}%
to improve readability, performance graphs for these less representative cases are deferred to Appendix~\ref{sec:extra-evaluation}.
\end{arxiv}

Our algorithm outperforms PMwCAS for list sizes 5 and 50 by 2.6$\times$ and 2.2$\times$ on average, respectively. 
This shows that under high and moderate contention, our algorithm's faster \mcas{} operations (due to the double effect of lower complexity and lower helping ratio) compensate for its slower read operations (due to the extra level of indirection), even in read-heavy workloads. In the low contention case (list size 500), PMwCAS outperforms our algorithm at low thread counts and is outperformed at high thread counts. On average, PMwCAS outperforms our algorithm by 1.2$\times$. Under low contention, operations have a low probability to conflict on the same element and thus the lower read complexity of PMwCAS has a stronger impact on performance than the lower \mcas{} complexity of our algorithm.


\subsection{B+-tree Benchmark}
In this benchmark we operate on a B+-tree which supports search and update (insert and delete) operations. Insertions and deletions use $k$-\cas{}, where $k$ may vary, e.g., depending on whether the operation led to nodes being split or merged.

Similar to the previous benchmark, we initialize the B+-tree with a configurable number of entries; threads then select operations and values at random. We perform the benchmark with 80\% reads and three initial tree sizes (16, 512, and 4000000). As for the previous benchmark, performance graphs for the 50\%, 98\% and 100\% reads cases are shown in
\begin{camera}
the full version of our paper~\cite{longversion}.
\end{camera}\begin{arxiv}
Appendix~\ref{sec:extra-evaluation}.
\end{arxiv} 
As before, the initial size of the tree is half of the key range. Results are shown in the bottom row of Figure~\ref{fig:dll}.

We observe a similar behavior to the previous benchmark. Our algorithm outperforms PMwCAS under high and medium contention (because it performs fewer \cas{}es and triggers less helping) and is slightly outperformed under low contention (where helping no longer plays a major role).

\section{Related Work}
\label{sec:related-work}

\xparagraph{Lock- and wait-free implementations of \mcas{}.} 
Our algorithm shares similarities with previous work~\cite{harris02,wang18}: as has become standard practice, it uses operation descriptors and a three-phase design (locking, status-change and unlocking). However, our algorithm introduces key differences with respect to previous work: it defers the unlocking phase and combines it with the reclamation of descriptors, without compromising correctness. This deferment has a triple beneficial effect on complexity: (1) it removes $k$ \cas{}es from the critical path, (2) it allows these \cas{}es to be amortized across several operations, and (3) it removes the onus of ABA-prevention from the locking phase, thus shaving off $k$ further \cas{}es from the latter.

Table~\ref{comparison} summarizes the differences between our algorithm and  existing non-blocking \mcas{} implementations, while the detailed treatment of each of the numerous prior efforts is deferred to
\begin{camera} 
the full version of our paper~\cite{longversion}.
\end{camera}\begin{arxiv}
Appendix~\ref{sec:related-non-blocking-mcas}. 
\end{arxiv} 
The results in Table~\ref{comparison} reflect the number of \cas{}es per \mcas{} operation required for correctness by each algorithm in the common uncontended case.
We note that previous \mcas{} implementations perform descriptor cleanup immediately after applying \mcas{}, and it is not clear how to separate cleanup from these algorithms while preserving correctness.
If we take the cleanup cost into consideration for our algorithm as well, its theoretical (worst-case) complexity becomes $2k+1$, the same as some of the previous work.
As our experiments in Section~\ref{sec:eval} demonstrate,
however, the number of \cas{}es in the cleanup phase is negligible in practice. Furthermore, we highlight the fact that unlike most previous work, including the one that employs $2k+1$ \cas{}es, readers in our case do not write into the shared memory in the common case, even when cleanup is considered.

\begin{table}
\centering
\caption{Comparison of non-blocking \mcas{} implementations in terms of the number of \cas{} instructions required, whether readers perform writes to shared memory or expensive atomic instructions, and the number of persistent fences (all per $k$-word \mcas{}, in the uncontended case). }
\begin{tabular}{@{}llll@{}}
  \toprule
         & \cas{}es & Readers write & P. fences \tabularnewline
  \midrule
  Israeli and Rappoport~\cite{israeli94} & $3k+2$ & Yes & N/A \tabularnewline
  Anderson and Moir~\cite{anderson95} & $3k+2$ & Yes & N/A \tabularnewline
  Moir~\cite{moir97} & $3k+4$ & Yes & N/A \tabularnewline
  Harris et al.~\cite{harris02} & $3k+1$ & No & N/A \tabularnewline
  Ha and Tsigas~\cite{ha03,ha04} & $2k+2$ & Yes & N/A \tabularnewline
  Attiya and Hillel~\cite{attiya11} & $6k+2$ & N/A & N/A \tabularnewline
  Sundell~\cite{sundell11} & $2k+1$ & Yes & N/A\tabularnewline
  Feldman et al.~\cite{feldman15} & $3k-1$ & Yes & N/A \tabularnewline
  Wang et al.~\cite{wang18} (volatile) & $3k+1$ & No & N/A \tabularnewline
  Wang et al.~\cite{wang18} (persistent) & $5k+1$ & No & $2k+1$ \tabularnewline
  \midrule\midrule
  {\bf Our algorithm} & $k+1$ & No & 2 \tabularnewline
  \bottomrule
\end{tabular}
\label{comparison}
\end{table}

\xparagraph{General techniques.}
Transactional memory (TM)~\cite{herlihy93a,shavit95} can be seen as the most general approach to  providing atomic access to multiple objects. It allows a block of code to be designated as a transaction and thus executed atomically, with respect to other transactions. Thus, TM is strictly more general than \mcas{}. This generality comes at a cost: software implementations of transactional memory (STM) have prohibitive performance overheads, whereas hardware support (HTM) is subject to spurious aborts and thus only provides ``best-effort'' guarantees. Prior work on nonblocking STMs~\cite{Fraser04, MaratheM08} share goals similar to our work; namely reduction of overheads in the
critical path.  However, these works (i) either employ $k$ extra cleanup CASes~\cite{Fraser04} on the critical path, incurring
precisely the overheads we avoid in our work, or (ii) employ a vastly more complex ``stealing'' framework to
avoid overheads from the critical path~\cite{MaratheM08}.

\begin{camera} 
\end{camera}\begin{arxiv}
As any concurrent object, \mcas{} can be implemented using a universal construction~\cite{herlihy93}, but such an implementation is not disjoint-access-parallel and has high overhead.
\end{arxiv}

\xparagraph{Prior Work on Persistent \mcas{}.}
Pavlovic et al.~\cite{pavlovic18} provide an implementation of \mcas{} for persistent memory which differs from ours in the progress guarantee (theirs is blocking) and hardware assumpions (theirs uses HTM). 
\begin{camera} 
\end{camera}\begin{arxiv} 
In their algorithm, a transaction is used to atomically verify expected values and acquire ownership of all target locations. In case of success, the new values are written non-transactionally. Reads that encounter a location owned by an \mcas{} operation block until the location is no longer owned.
\end{arxiv}

Wang et al.~\cite{arulraj18,wang18} introduce the first lock-free persistent implementation, based on the algorithm of Harris et al.~\cite{harris02}. The main differences with respect to our algorithm are outlined in Table~\ref{comparison}. This algorithm uses a per-word dirty flag to indicate that the word is not yet guaranteed to be written to persistent memory. Operations encountering a set dirty flag will persist the associated word and then unset the flag. This technique avoids unnecessary persistent flushes, but uses 2 extra \cas{} instructions per target location in order to manipulate the dirty flag. \begin{camera}
\end{camera}\begin{arxiv}
In total, this algorithm uses $5k+1$ \cas{} instructions for a $k$-word \mcas{} in the uncontended case. Their implementation does not use explicit persistent fences; instead, it relies on the \cas{} instructions that are already required to unset the dirty flag to also enforce ordering among write backs~\cite{intel}. Their original algorithm uses $2k+1$ such ``\cas{}-fences'', but we believe it can be modified to only require 3 persistent fences.
\end{arxiv}

In our work we use the recent durable linearizability correctness condition~\cite{izraelevitz16}, which assumes a full-system crash-recovery model, but other models of persistent memory can be explored in this context~\cite{aguilera2003strict, BerryhillGT15, ConditNFILBC09, GuerraouiL04, PelleyCW15}.

\section{Conclusion}
\label{sec:conclusion}


Atomic multi-word primitives significantly simplify concurrent algorithm design, but existing implementations have high overhead. In this paper, we propose a simple and efficient lock-free algorithm for multi-word compare-and-swap, designed for both volatile and persistent memory. 
The complimentary lower bound shows that the complexity of our algorithm, as measured in the
number of \cas{}es in the uncontended case, is nearly optimal.


\bibliography{bib/main}

\begin{thebibliography}{10}

\bibitem{AfekDM11}
Yehuda Afek, Dave Dice, and Adam Morrison.
\newblock Cache index-aware memory allocation.
\newblock In {\em Proceedings of the International Symposium on Memory
  Management (ISMM)}, page 55–64. Association for Computing Machinery, 2011.

\bibitem{aguilera2003strict}
Marcos~K Aguilera and Svend Fr{\o}lund.
\newblock Strict linearizability and the power of aborting.
\newblock Technical Report HPL-2003-241, HP Labs, 2003.

\bibitem{alistarh17}
Dan Alistarh, William~M. Leiserson, Alexander Matveev, and Nir Shavit.
\newblock {Forkscan: Conservative Memory Reclamation for Modern Operating
  Systems}.
\newblock In {\em Proceedings of the Twelfth European Conference on Computer
  Systems, EuroSys 2017}, pages 483--498, 2017.

\bibitem{anderson95}
James~H. Anderson and Mark Moir.
\newblock {Universal Constructions for Multi-object Operations}.
\newblock In {\em {14th Annual ACM Symposium on Principles of Distributed
  Computing}}, pages 184--193, 1995.

\bibitem{anderson97}
James~H. Anderson, Srikanth Ramamurthy, and Rohit Jain.
\newblock {Implementing Wait-free Objects on Priority-based Systems}.
\newblock In {\em {16th Annual ACM Symposium on Principles of Distributed
  Computing}}, pages 229--238, 1997.

\bibitem{arulraj18}
Joy Arulraj, Justin Levandoski, Umar~Farooq Minhas, and Per-Ake Larson.
\newblock {BzTree: A High-Performance Latch-free Range Index for Non-Volatile
  Memory}.
\newblock In {\em {44th International Conference on Very Large Data Bases}},
  2018.

\bibitem{ascylibcode}
{ASCYLIB}, a concurrent-search data-structure library with over 40
  implementations of linked lists, hash tables, skip lists, binary search
  trees, queues, and stacks.
\newblock \url{https://github.com/LPD-EPFL/ASCYLIB}, 2018.

\bibitem{attiya11}
Hagit Attiya and Eshcar Hillel.
\newblock Highly concurrent multi-word synchronization.
\newblock {\em Theor. Comput. Sci.}, 412(12-14):1243--1262, 2011.

\bibitem{BerryhillGT15}
Ryan Berryhill, Wojciech~M. Golab, and Mahesh Tripunitara.
\newblock Robust shared objects for non-volatile main memory.
\newblock In {\em 19th International Conference on Principles of Distributed
  Systems, {OPODIS} 2015}, pages 20:1--20:17, 2015.

\bibitem{braginsky12}
Anastasia Braginsky and Erez Petrank.
\newblock {A Lock-free B+Tree}.
\newblock In {\em Proceedings of the Twenty-fourth Annual ACM Symposium on
  Parallelism in Algorithms and Architectures}, pages 58--67, 2012.

\bibitem{brown13}
Trevor Brown, Faith Ellen, and Eric Ruppert.
\newblock {Pragmatic Primitives for Non-blocking Data Structures}.
\newblock In {\em {ACM} Symposium on Principles of Distributed Computing,
  {PODC} '13}, pages 13--22, 2013.

\bibitem{Brown15}
Trevor~Alexander Brown.
\newblock {Reclaiming Memory for Lock-Free Data Structures: There Has to Be a
  Better Way}.
\newblock In {\em Proceedings of the 2015 ACM Symposium on Principles of
  Distributed Computing}, pages 261--270, 2015.

\bibitem{bztreecode}
Bztree: a high-performance latch-free range index for non-volatile memory.
\newblock \url{https://github.com/wangtzh/bztree}, 2019.

\bibitem{CepedaCLLWG19}
Diego Cepeda, Sakib Chowdhury, Nan Li, Raphael Lopez, Xinzhe Wang, and Wojciech
  Golab.
\newblock Toward linearizability testing for multi-word persistent
  synchronization primitives.
\newblock In {\em 23rd International Conference on Principles of Distributed
  Systems, {OPODIS} 2019}, volume 153, pages 19:1--19:17, 2019.

\bibitem{cohen18}
Nachshon Cohen, Rachid Guerraoui, and Igor Zablotchi.
\newblock {The Inherent Cost of Remembering Consistently}.
\newblock In {\em Proceedings of the 30th {ACM} Symposium on Parallelism in
  Algorithms and Architectures, {SPAA} 2018}, 2018.

\bibitem{ConditNFILBC09}
Jeremy Condit, Edmund~B. Nightingale, Christopher Frost, Engin Ipek,
  Benjamin~C. Lee, Doug Burger, and Derrick Coetzee.
\newblock Better {I/O} through byte-addressable, persistent memory.
\newblock In {\em Proceedings of the 22nd {ACM} Symposium on Operating Systems
  Principles 2009, {SOSP} 2009}, pages 133--146, 2009.

\bibitem{david18}
Tudor David, Aleksandar Dragojevic, Rachid Guerraoui, and Igor Zablotchi.
\newblock {Log-Free Concurrent Data Structures}.
\newblock In {\em 2018 {USENIX} Annual Technical Conference, {USENIX} {ATC}
  2018}, 2018.

\bibitem{DavidGT13}
Tudor David, Rachid Guerraoui, and Vasileios Trigonakis.
\newblock Everything you always wanted to know about synchronization but were
  afraid to ask.
\newblock In {\em {ACM} {SIGOPS} 24th Symposium on Operating Systems
  Principles, {SOSP} '13, Farmington, PA, USA, November 3-6, 2013}, pages
  33--48, 2013.

\bibitem{detlefs00}
David Detlefs, Christine~H. Flood, Alex Garthwaite, Paul Martin, Nir Shavit,
  and Guy~L. Steele, Jr.
\newblock {Even Better DCAS-Based Concurrent Deques}.
\newblock In {\em {14th International Conference on Distributed Computing}},
  pages 59--73, 2000.

\bibitem{DHK16}
Dave Dice, Maurice Herlihy, and Alex Kogan.
\newblock Fast non-intrusive memory reclamation for highly-concurrent data
  structures.
\newblock In {\em Proceedings of the 2016 ACM SIGPLAN International Symposium
  on Memory Management}, pages 36--45, 2016.

\bibitem{doherty04}
Simon Doherty, David~L. Detlefs, Lindsay Groves, Christine~H. Flood, Victor
  Luchangco, Paul~A. Martin, Mark Moir, Nir Shavit, and Guy~L. Steele, Jr.
\newblock {DCAS is Not a Silver Bullet for Nonblocking Algorithm Design}.
\newblock In {\em Proceedings of the Sixteenth Annual ACM Symposium on
  Parallelism in Algorithms and Architectures}, pages 216--224, 2004.

\bibitem{feldman15}
Steven~D. Feldman, Pierre LaBorde, and Damian Dechev.
\newblock {A Wait-Free Multi-Word Compare-and-Swap Operation}.
\newblock {\em International Journal of Parallel Programming}, 43(4):572--596,
  2015.

\bibitem{Fraser04}
Keir Fraser.
\newblock {\em Practical lock-freedom}.
\newblock PhD thesis, University of Cambridge, {UK}, 2004.

\bibitem{FriedmanHMP18}
Michal Friedman, Maurice Herlihy, Virendra~J. Marathe, and Erez Petrank.
\newblock A persistent lock-free queue for non-volatile memory.
\newblock In {\em Proceedings of the 23rd {ACM} {SIGPLAN} Symposium on
  Principles and Practice of Parallel Programming, {PPoPP} 2018}, pages 28--40,
  2018.

\bibitem{greenwald99}
Michael Greenwald.
\newblock Non-blocking synchronization and system design. {\it ph.d. thesis,
  stanford university}, 1999.

\bibitem{greenwald02}
Michael Greenwald.
\newblock Two-handed emulation: how to build non-blocking implementation of
  complex data-structures using {DCAS}.
\newblock In {\em 21st Annual {ACM} Symposium on Principles of Distributed
  Computing}, pages 260--269, 2002.

\bibitem{longversion}
Rachid Guerraoui, Alex Kogan, Virendra~J. Marathe, and Igor Zablotchi.
\newblock Efficient multi-word compare and swap.
\newblock {\em ArXiv preprint arXiv:???}, 2020.
\newblock URL: \url{http://arxiv.org/???}, \href {http://arxiv.org/abs/???}
  {\path{arXiv:???}}

\bibitem{GuerraouiL04}
Rachid Guerraoui and Ron~R. Levy.
\newblock {Robust Emulations of Shared Memory in a Crash-Recovery Model}.
\newblock In {\em 24th International Conference on Distributed Computing
  Systems {(ICDCS} 2004)}, pages 400--407, 2004.

\bibitem{ha03}
Phuong~Hoai Ha and Philippas Tsigas.
\newblock {Reactive Multi-Word Synchronization for Multiprocessors}.
\newblock In {\em 12th International Conference on Parallel Architectures and
  Compilation Techniques {(PACT} 2003)}, pages 184--193, 2003.

\bibitem{ha04}
Phuong~Hoai Ha and Philippas Tsigas.
\newblock {Reactive Multi-word Synchronization for Multiprocessors}.
\newblock {\em J. Instruction-Level Parallelism}, 6, 2004.

\bibitem{harris10}
Tim Harris, James~R. Larus, and Ravi Rajwar.
\newblock {\em {Transactional Memory: 2nd Edition}}.
\newblock {Morgan \& Claypool}, 2010.

\bibitem{harris03}
Timothy~L. Harris and Keir Fraser.
\newblock {Language support for lightweight transactions}.
\newblock In {\em Proceedings of the 2003 {ACM} {SIGPLAN} Conference on
  Object-Oriented Programming Systems, Languages and Applications}, pages
  388--402, 2003.

\bibitem{harris02}
Timothy~L. Harris, Keir Fraser, and Ian~A. Pratt.
\newblock {A Practical Multi-word Compare-and-Swap Operation}.
\newblock In {\em 16th International Conference on Distributed Computing},
  pages 265--279, 2002.

\bibitem{herlihy93}
Maurice Herlihy.
\newblock A methodology for implementing highly concurrent data objects.
\newblock {\em {ACM Transactions on Programming Languages and Systems}},
  15(5):745--770, November 1993.

\bibitem{herlihy03}
Maurice Herlihy, Victor Luchangco, Mark Moir, and William~N. Scherer, III.
\newblock {Software Transactional Memory for Dynamic-sized Data Structures}.
\newblock In {\em Proceedings of the Twenty-second Annual Symposium on
  Principles of Distributed Computing}, pages 92--101, 2003.

\bibitem{herlihy93a}
Maurice Herlihy and J.~Eliot~B. Moss.
\newblock {Transactional Memory: Architectural Support for Lock-free Data
  Structures}.
\newblock In {\em {20th Annual International Symposium on Computer
  Architecture}}, pages 289--300, 1993.

\bibitem{HW+91}
Maurice Herlihy and Jeannette~M Wing.
\newblock {Linearizability: A correctness condition for concurrent objects}.
\newblock {\em ACM Transactions on Programming Languages and Systems},
  12(3):463--492, 1990.

\bibitem{intel}
Intel.
\newblock {Intel\textsuperscript{\textregistered} 64 and IA-32 Architectures
  Software Developer's Manual Combined}.
\newblock
  \url{https://software.intel.com/sites/default/files/managed/39/c5/325462-sdm-vol-1-2abcd-3abcd.pdf},
  2018.

\bibitem{israeli94}
Amos Israeli and Lihu Rappoport.
\newblock {Disjoint-Access-Parallel Implementations of Strong Shared Memory
  Primitives}.
\newblock In {\em Proceedings of the Thirteenth Annual {ACM} Symposium on
  Principles of Distributed Computing}, pages 151--160, 1994.

\bibitem{izraelevitz16}
Joseph Izraelevitz, Hammurabi Mendes, and Michael~L. Scott.
\newblock Linearizability of persistent memory objects under a
  full-system-crash failure model.
\newblock In {\em Distributed Computing - 30th International Symposium, {DISC}
  2016}, pages 313--327, 2016.

\bibitem{levandoski13}
Justin~J. Levandoski, David~B. Lomet, and Sudipta Sengupta.
\newblock {The Bw-Tree: A B-tree for New Hardware Platforms}.
\newblock In {\em Proceedings of the 2013 IEEE International Conference on Data
  Engineering (ICDE 2013)}, pages 302--313, 2013.

\bibitem{luchangco03}
Victor Luchangco, Mark Moir, and Nir Shavit.
\newblock {Nonblocking k-compare-single-swap}.
\newblock In {\em {SPAA} 2003: Proceedings of the Fifteenth Annual {ACM}
  Symposium on Parallelism in Algorithms and Architectures}, pages 314--323,
  2003.

\bibitem{MaratheM08}
Virendra~J. Marathe and Mark Moir.
\newblock Toward high performance nonblocking software transactional memory.
\newblock In {\em Proceedings of the 13th {ACM} {SIGPLAN} Symposium on
  Principles and Practice of Parallel Programming, {PPOPP} 2008, Salt Lake
  City, UT, USA, February 20-23, 2008}, pages 227--236. {ACM}, 2008.

\bibitem{McKenney17}
Paul~E. McKenney.
\newblock Is parallel programming hard, and, if so, what can you do about it?,
  2017.

\bibitem{MS98}
Paul~E. McKenney and John~D. Slingwine.
\newblock {Read-Copy Update: Using Execution History to Solve Concurrency
  Problems}.
\newblock In {\em {Parallel and Distributed Computing and Systems}}, 1998.

\bibitem{Michael04}
Maged~M. Michael.
\newblock Hazard pointers: Safe memory reclamation for lock-free objects.
\newblock {\em IEEE Trans. Parallel Distrib. Syst.}, 15(6):491--504, 2004.

\bibitem{moir97}
Mark Moir.
\newblock {Transparent Support for Wait-Free Transactions}.
\newblock In {\em {11th International Workshop on Distributed Algorithms}},
  pages 305--319, 1997.

\bibitem{natarajan14}
Aravind Natarajan and Neeraj Mittal.
\newblock {Fast Concurrent Lock-free Binary Search Trees}.
\newblock In {\em Proceedings of the 19th ACM SIGPLAN Symposium on Principles
  and Practice of Parallel Programming}, pages 317--328, 2014.

\bibitem{openmp}
The {OpenMP API} specification for parallel programming.
\newblock \url{https://www.openmp.org/}, 2019.

\bibitem{pavlovic18}
Matej Pavlovic, Alex Kogan, Virendra~J. Marathe, and Tim Harris.
\newblock {Persistent Multi-Word Compare-and-Swap}.
\newblock In {\em {ACM Synposium on Principles of Distributed Computing}},
  2018.

\bibitem{PelleyCW15}
Steven Pelley, Peter~M. Chen, and Thomas~F. Wenisch.
\newblock Memory persistency: Semantics for byte-addressable nonvolatile memory
  technologies.
\newblock {\em {IEEE} Micro}, 35(3):125--131, 2015.

\bibitem{pibencode}
Benchmarking framework for index structures on persistent memory.
\newblock \url{https://github.com/wangtzh/pibench}, 2019.

\bibitem{pmwcascode}
Persistent multi-word compare-and-swap ({PMwCAS}) for {NVRAM}.
\newblock \url{https://github.com/microsoft/pmwcas}, 2019.

\bibitem{poter18}
Manuel P{\"{o}}ter and Jesper~Larsson Tr{\"{a}}ff.
\newblock \emph{Stamp-it}, amortized constant-time memory reclamation in
  comparison to five other schemes.
\newblock In {\em Proceedings of the 23rd {ACM} {SIGPLAN} Symposium on
  Principles and Practice of Parallel Programming, PPoPP 2018}, pages 413--414,
  2018.

\bibitem{shavit95}
Nir Shavit and Dan Touitou.
\newblock {Software Transactional Memory}.
\newblock In {\em {14th Annual ACM Symposium on Principles of Distributed
  Computing}}, pages 204--213, 1995.

\bibitem{sundell11}
H{\aa}kan Sundell.
\newblock {Wait-Free Multi-Word Compare-and-Swap Using Greedy Helping and
  Grabbing}.
\newblock {\em International Journal of Parallel Programming}, 39(6):694--716,
  2011.

\bibitem{timnat15}
Shahar Timnat, Maurice Herlihy, and Erez Petrank.
\newblock {A Practical Transactional Memory Interface}.
\newblock In {\em Euro-Par 2015: Parallel Processing - 21st International
  Conference on Parallel and Distributed Computing}, pages 387--401, 2015.

\bibitem{wang18}
Tianzheng Wang, Justin Levandoski, and Per-Ake Larson.
\newblock {Easy Lock-Free Indexing in Non-Volatile Memory}.
\newblock In {\em {34th IEEE International Conference on Data Engineering}},
  2018.

\bibitem{wen18}
Haosen Wen, Joseph Izraelevitz, Wentao Cai, H.~Alan Beadle, and Michael~L.
  Scott.
\newblock Interval-based memory reclamation.
\newblock In {\em Proceedings of the 23rd {ACM} {SIGPLAN} Symposium on
  Principles and Practice of Parallel Programming}, pages 1--13, 2018.

\end{thebibliography}

\begin{camera}
\end{camera}\begin{arxiv}
\appendix
\newpage
\section{APPENDIX}

\subsection{Replacing RDCSS with CAS in Harris et al. algorithm leads to ABA}\label{sec:aba-harris}

Consider two memory locations $a_1$ and $a_2$ with initial values $v_1$ and $v_2$ respectively. Consider two $2$-\cas{} operations $op$ and $op'$ which operate on $a_1$ and $a_2$. $op$ has old values $v_1$ and $v_2$ and new values $v_1'$ and $v_2'$, respectively; $op'$ has old values $v_1'$ and $v_2'$ and new values $v_1$ and $v_2$, respectively. Let $D$ be $op$'s descriptor.

\begin{enumerate}
    \item $op$ executes solo and performs the \cas{} to make $a_1$ point to $D$, then pauses immediately before the \cas{} to acquire $a_2$.
    \item $op'$ executes solo: it first helps $op$ complete, changing the values of $a_1$ and $a_2$ to $v_1'$ and $v_2'$ respectively; then $op'$ performs its own changes, modifying $a_1$'s and $a_2$'s values back to $v_1$ and $v_2$, respectively. 
    \item $op$ resumes, successfully acquires $a_2$, performs the status-change \cas{} on $D$, then performs unlocking \cas{}es on $a_1$ and $a_2$. The \cas{} on $a_1$ will fail, and $a_1$'s value will remain $v_1$. The \cas{} on $a_2$ will succeed, changing its value to $v_2'$.
    \item The values of $a_1$ and $a_2$ are now incompatible with any linearization of $op$ and $op'$.
\end{enumerate}

\subsection{Correctness of Volatile \texorpdfstring{$k+1$}{k+1} algorithm}
\label{sec:appendix-correctness}
\label{sec:k+1-mcas-correctness}
In this section we argue that our \mcas{} algorithm is linearizable and lock-free. We give preliminary invariants before showing the main results. 

\begin{lemma}\label{finalized-status}
    Once an \mcas{} descriptor is finalized, its status never changes again. \normalfont{The status can only be modified through the \cas{} at Line~\ref{mcas-finalize}, whose expected value is \ttactive. If the \cas{} succeeds, the new value of the status can only be \successful\ or \failed, thus any subsequent attempt to change the status will fail.}
\end{lemma}

\begin{lemma}
    An \mcas{} descriptor is finalized by at most one thread. \normalfont{This follows from the fact that a descriptor is finalized through a \cas{} and the fact that an \mcas{} descriptor cannot change status after being finalized (Lemma~\ref{finalized-status})}.
\end{lemma}

\begin{lemma}
    If at least one thread attempts to finalize a descriptor $d$, some thread will successfully finalize $d$. \normalfont{The initial status of a descriptor is \ttactive. Any thread attempting to finalize a descriptor does so through the CAS at Line~\ref{mcas-finalize}, with expected value \ttactive. Thus, at least one CAS finds the status to be \ttactive\ and successfully changes it.}
\end{lemma}

\begin{lemma}\label{finalized-successful}
    An \mcas{} descriptor $d$ is finalized as successful only if some thread observed all target locations of $d$ to be acquired by $d$. \normalfont{This is because the status is changed to successful only if the \texttt{success} variable is true at Line~\ref{mcas-finalize}. This only happens if some thread completed the for-loop over all of $d$'s {\worddescriptor}s without exiting the loop at Line~\ref{for-break}. The only two ways for a thread to move to the next \worddescriptor\ in the loop is if the thread sees the current target location was already acquired by $d$ (Line~\ref{disregarded}) or if the thread successfully acquired the current target location for $d$ (Line~\ref{mcas-acquire}). In both cases the thread observed the target location to be acquired by $d$.}
\end{lemma}
 
\begin{lemma}\label{finalized-failed}
    An \mcas{} descriptor $d$ is finalized as failed only if some thread observed a target location of $d$ to contain a different value than its expected value in $d$. \normalfont{This is because the only way for the status to be changed to failed is if the \texttt{success} variable is false. This only happens if some thread observed the current value of a target location is different from its expected value in Line~\ref{doomed-start}.}
\end{lemma}

\begin{lemma}\label{lemma-acquired-forever}
    After a location $l$ becomes acquired by some operation $op$, $l$ will never become un-acquired again. \normalfont{This is because the only instruction that modifies a location $l$ is the acquire \cas{} at Line~\ref{mcas-acquire}.}
\end{lemma}

We say that an operation $op_1$ {\em helps} an \mcas{} operation $op_2$ if $op_1$ calls \mcas{} with $op_2$'s descriptor in Line~\ref{mcas-help}. 

\begin{lemma}\label{lemma-change-owner}
    After a location $l$ becomes acquired by some operation $op$, no operation $op' \ne op$ will acquire $l$ before $op$ becomes finalized. \normalfont{Assume by contradiction that $op'$ acquires $l$ after $op$ acquires $l$ and while $op$ is still active. Consider $op'$ last call to \readInternal\ (Line~\ref{mcas-readinternal}) before the successful acquisition of $l$. During this call, $op'$ must have observed that $l$ is owned by $op$ (otherwise; if $op$ had acquired $l$ after the call to \readInternal, the acquisition \cas{} would have failed). Moreover, $op$ was active during that call to \readInternal\ by $op'$. Thus, $op'$ helped $op$ before returning from \readInternal, finalizing $op$ in the process. Thus $op$ cannot be active at the time of the acquisition, a contradiction.}
\end{lemma}

We say that a location $l$ is \textit{re-acquired} by operation $op$ at time $t$ if (1) $l$ becomes acquired by $op$ at time $t$, (2) there exists time $t' < t$ such that $l$ became acquired by $op' \ne op$ at time $t'$, and (3) there exists time $t'' < t'$ such that $l$ became acquired by $op$ at time $t''$.

\begin{lemma}\label{lemma-no-reacquisition}
    A location cannot be re-acquired. 
    \normalfont{ Assume the contrary and let $t$ be the earliest time when any location is re-acquired in a given execution $E$. Let $l$ be that location and $op$ be the operation re-acquiring it. This means that $l$ became acquired by $op$ at some time $t''$, then became acquired by some $op'\ne op$ at time $t' > t$ and then later became acquired by $op$ at time $t > t'$ (the times in this lemma are represented in the figure below, for convenience). By Lemma~\ref{lemma-change-owner}, $op$ must have become finalized at some time $t_f < t'$ ($t_f$ is unique by Lemma~\ref{finalized-status}).
    
    Now consider the thread $T$ which acquires $l$ on behalf of $op$ at time $t$. $T$ does so through the \cas{} at line~\ref{mcas-acquire}. Since $op$ becomes finalized at time $t_f$, $T$ must have performed the status check at line~\ref{status-check} at some time $t_s < t_f$ (otherwise $T$ would have exited from the for loop without acquiring $l$). Let $t_r < t_s$ be the last time when $T$ performed \readInternal{} (line~\ref{mcas-readinternal}) before $t_s$. Note that $t_r < t''$, otherwise $T$ would have seen $l$ as already acquired by $op$ at line~\ref{disregarded} and continued without attempting to acquire $l$.  
    
    Let $\langle c,v \rangle$ be the return value of the \readInternal{} call by $T$ at $t_r$; this means that $l$'s value was $c$ at some time before $t_r$. Since $T$ successfully performs the CAS at line~\ref{mcas-acquire} at time $t$, the value of $l$ must also be $c$ immediately before $t$. However, the acquisition of $l$ at time $t'' > t_r$ changes the value of $l$ from $c$. Therefore, it must be the case that some thread changes the value of $l$ back to $c$ at some time $t_c$ between $t''$ and $t$. Note that $c$ must be a word descriptor (due to Lemma~\ref{lemma-acquired-forever}). Since word descriptors are unique, they uniquely identify their parent operations. Therefore, $l$ must have been owned by some operation $op''$ before $t_r$ and again at $t_c$; this means that $op''$ \textit{re-acquired} $l$ at time $t_c$, contradicting our choice of $t$ as the earliest re-acquisition time.}
    
\end{lemma}

\noindent
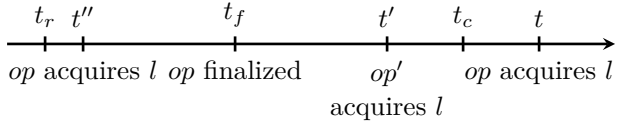
\begin{tikzpicture}
\usetikzlibrary{calc}

\coordinate (start) at (-4,0);
\coordinate (end) at (4,0);
\draw [line width=1pt, -stealth] (start) -- (end);

\coordinate (s0) at (-3,0);
\coordinate (t00) at ($(s0)+(0,0.1)$);
\coordinate (t01) at ($(s0)-(0,0.1)$);
\coordinate (s1) at (1,0);
\coordinate (t10) at ($(s1)+(0,0.1)$);
\coordinate (t11) at ($(s1)-(0,0.1)$);
\coordinate (s2) at (3,0);
\coordinate (t20) at ($(s2)+(0,0.1)$);
\coordinate (t21) at ($(s2)-(0,0.1)$);
\coordinate (s3) at (-1,0);
\coordinate (t30) at ($(s3)+(0,0.1)$);
\coordinate (t31) at ($(s3)-(0,0.1)$);
\coordinate (s4) at (-3.5,0);
\coordinate (t40) at ($(s4)+(0,0.1)$);
\coordinate (t41) at ($(s4)-(0,0.1)$);
\coordinate (s5) at (2,0);
\coordinate (t50) at ($(s5)+(0,0.1)$);
\coordinate (t51) at ($(s5)-(0,0.1)$);

\draw [line width=1pt] (t00) -- (t01);
\node [anchor=south] at (t00.north) {$t''$};

\draw [line width=1pt] (t10) -- (t11);
\node [anchor=south] at (t10.north) {$t'$};

\draw [line width=1pt] (t20) -- (t21);
\node [anchor=south] at (t20.north) {$t$};

\draw [line width=1pt] (t30) -- (t31);
\node [anchor=south] at (t30.north) {$t_f$};

\draw [line width=1pt] (t40) -- (t41);
\node [anchor=south] at (t40.north) {$t_r$};

\draw [line width=1pt] (t50) -- (t51);
\node [anchor=south] at (t50.north) {$t_c$};

\node [anchor=north, align=center, text width=2cm] at (t01.south) {
$op$ acquires $l$
};

\node [anchor=north, align=center, text width=2cm] at (t11.south) {
$op'$ acquires $l$
};

\node [anchor=north, align=center, text width=2cm] at (t21.south) {
$op$ acquires $l$
};

\node [anchor=north, align=center, text width=2cm] at (t31.south) {
$op$ finalized
};

\end{tikzpicture}

\begin{lemma}
    A location $l$ cannot be acquired by operation $op$ after $op$ is finalized as successful.
    \normalfont{This follows from Lemmas~\ref{finalized-successful} and \ref{lemma-no-reacquisition}.}
\end{lemma}


\begin{lemma}\label{lemma-higher-highest}
    If $op_1$ helps $op_2$, then either $op_2$ highest acquired location is higher than $op_1$'s highest acquired location, or $op_1$ has not acquired any locations. \normalfont{If $op_1$ is a read operation, the statement is trivially true. Assume now that $op_1$ is an \mcas{} operation that helps $op_2$ and that $op_1$'s highest acquired location is higher than $op_2$'s highest acquired location ($\star$). Since $op_1$ helps $op_2$, $op_1$ has observed one of its target locations $l$ to be already acquired by $op_2$. But since $op_1$ iterates over locations in increasing order, $l$ must be higher than $op_1$'s highest acquired location. This contradicts $\star$.}
\end{lemma}

We define the {\em helping graph at time $t$}, $H(t)$, as follows. The vertices of $H(t)$ are the ongoing operations at time $t$. There is an edge from $op_1$ to $op_2$ if $op_1$ is helping $op_2$ at $t$. We define the {\em call depth} of an operation $op$ at time $t$ to be the length of the longest path starting from $op$ in $H(t)$.
    
    
\begin{lemma}\label{lemma-call-depth}
    For any operation $op$ and any time $t$, the call depth of $op$ at $t$ is finite. \normalfont{Assume the contrary. Since each thread can have at most one ongoing operation at $t$, $H(t)$ has a finite vertex set. Let $op$ be an operation and $t$ be a time such that $op$ has an infinite call depth at $t$. Then, $H(t)$ must contain a cycle. This is a contradiction: if the cycle contains an operation $op_0$ that has no acquired locations, then $op_0$'s predecessor in the cycle cannot be helping it; if the cycle does not contain such an operation, then by traversing the cycle we would find operations with strictly increasing highest acquired locations (Lemma~\ref{lemma-higher-highest}).}
\end{lemma}

Informally, Lemma~\ref{lemma-call-depth} says that while in our algorithm it is possible for operations to recursively help one another, the recursion depth is finite at any time, due to the sorting of memory locations.

We define the following predicates (recall that $n$ is the number of threads). Let $S(k)$: ``If there are $0 < k \leq n$ concurrent operations and at least one thread is taking steps and no operations are created, at least one operation will eventually return''. Let $P(k)$: ``If there are $0 < k \leq n$ concurrent operations and at least one thread is taking steps, at least one operation will eventually return''.

\begin{lemma}\label{sk}
    $S(k)$ is true for all $k$, $0 < k \leq n$. \normalfont{Assume the contrary. Pick an active thread $T$: $T$ is taking steps infinitely often, but no operations ever return. By Lemma~\ref{lemma-call-depth}, the call  depth of $T$ is finite, thus $T$ must be taking some backward branch infinitely often. If $T$ is taking the branch at Line~\ref{read-internal-retry} infinitely often, then \mcas{} operations are being finalized infinitely often (Line~\ref{read-internal-retry} is only executed if some operation was active at Line~\ref{read-internal-check}; but that same operation must be finalized by Line~\ref{read-internal-retry} due to the preceding \mcas{} call which returns only after the operation is finalized). This is a contradiction because we started with a finite number of \mcas{} operations and no operations are being created. If $T$ is taking the branch at Line~\ref{mcas-retry} infinitely often, then locations either (a) become acquired infinitely often or (b) change owners infinitely often. Both possibilities lead to a contradiction: (a) because there are a finite number of target locations of ongoing \mcas{} operation and locations never become unacquired and (b) because locations change owners only after operations become finalized, which would imply that operations become finalized infinitely often.}
\end{lemma}

\begin{lemma}\label{lemma-pk}
    $P(k)$ is true for all $k$, $0 < k \leq n$. \normalfont{Consider the case $k=n$. $P(k)$ is equivalent to $S(k)$ in this case (no operations can be created if there are already as many operations as threads), and thus true. Consider the case $k = n-1$. If some operation is eventually created, then eventually some operation will return, by $P(n)$. If no operation is ever created, then eventually some operation will return, by $S(n-1)$. We can continue in this manner with $k=n-2,...,1$, each time using either $P(k+1)$ or $S(k)$.}
\end{lemma}

\begin{lemma}
    Our implementation is lock-free. \normalfont{This follows immediately from Lemma~\ref{lemma-pk}.}
\end{lemma}

\begin{lemma}
    Linearization point of a failed \mcas{}. \normalfont{By Lemma~\ref{finalized-failed}, if descriptor $d$ is finalized as failed by thread $T$ at time $t_1$, then at time $t_0 < t_1$, $T$ has observed some target location $l$ to contain a different value than $l$'s expected value in $d$. We can take $t_0$ as the linearization point of the \mcas{}.}
\end{lemma}

\begin{lemma}
    Linearization point of a successful \mcas{}. \normalfont{By Lemma~\ref{finalized-successful}, if thread $T$ changes the status of descriptor $d$ to successful, then $T$ previously observed all of $d$'s target locations to be acquired by $d$. Thus, when changing the status of $d$ to successful, $T$ changes the logical values of all target locations, marking the linearization point.}
\end{lemma}

\begin{lemma}
    Linearization point of a read. \normalfont{The linearization point of a read is the last executed dereference instruction at Line~\ref{read-internal-first-read}.} 
\end{lemma}

\subsection{Persistent \mcas{} with \texorpdfstring{$k+1$}{k+1} \cas{} and 2 Persistent Fences}\label{app:persistent-mcas}
\bigskip
\begin{lstlisting}[caption={[Our \mcas{} algorithm for persistent memory.]Our \mcas{} algorithm for persistent memory. Commands in {\it italic} are related to memory reclamation, and \uline{underlined} commands are related to persistence.},label=list:algo-persistent,firstnumber=1,keywords={}]
read(void* address) { @\label{pread_begin}@
    @{\it epochStart();}@@\label{pread-epoch-start}@
    <content, value> = readInternal(address, NULL);@\label{pread-line2}@
    @{\it epochEnd();}@@\label{pread-epoch-end}@
    return value; } @\label{pread_end}@

MCAS(MCASDescriptor* desc) {
    @{\it epochStart();}@
    success = true;
    for wordDesc in desc->words {@\label{pmcas-first-phase-begin}@
retry_word:
        <content, value> = readInternal(wordDesc.address, desc); @\label{pmcas-readinternal}@
        // if this word already points to the right place, move on
        if (content == &wordDesc) continue;@\label{pdisregarded}@
        // if the expected value is different, the MCAS fails
        if (value != wordDesc.old) {@\label{pdoomed-start}@ success = false; break; @\label{pfor-break}@ }@\label{pdoomed-end}@
        if (desc->status != ACTIVE) break; @\label{pstatus-check}@
        // try to install the pointer to my descriptor; if failed,    retry
        if (!CAS(wordDesc.address, content, &wordDesc)) @\label{pmcas-acquire}@goto retry_word;@\label{pmcas-retry}@ }@\label{pmcas-first-phase-end}@
    @\uline{for wordDesc in desc->words \{ PERSISTENT\_FLUSH(wordDesc.address); \}}@
    @\uline{PERSISTENT\_FENCE();}\label{fence-acquire}@
    newStatus = success ? SUCCESSFUL : FAILED;
    if (CAS(&desc.status, ACTIVE, newStatus @\uline{| DirtyFlag}@)){@\label{pmcas-finalize}@
        // if I finalized this descriptor, mark it for reclamation
        @{\it retireForCleanup(desc);}@@\label{pretire-for-cleanup}@ }@\label{pmcas-finalize-end}@
    @\uline{PERSISTENT\_FLUSH(\&desc.status);}@
    @\uline{PERSISTENT\_FENCE();}\label{fence-finalize}@
    @\uline{parent->status = parent->status \& \textasciitilde{}DirtyFlag;}@ @\label{unset-dirty}@
    returnValue = (desc.status == SUCCESSFUL);
    @{\it epochEnd();}@
    return returnValue; }
\end{lstlisting}
\newpage
\begin{lstlisting}[caption={[The \readInternal\ auxiliary function, used by our \mcas{} algorithm for persistent memory.]The \readInternal\ auxiliary function, used by our \mcas{} algorithm for persistent memory. \uline{Underlined} commands are related to persistence.},label=list:readInternal-persistent,keywords={},firstnumber=last]
readInternal(void* addr, MCASDescriptor *self) {@\label{pr_i_begin}@
retry_read:                            
    val = *addr;@\label{pread-internal-first-read}@
    if (!isDescriptor(val))@\label{pread-internal-line4}@ then return <val,val>;@\label{pread-internal-line5}@
    else { // found a descriptor@\label{pread-internal-line6}@
        MCASDescriptor* parent = val->parent;@\label{pread-internal-line7}@
        if (parent != self) && parent->status == ACTIVE) {@\label{pread-internal-check}@
            MCAS(parent);@\label{pmcas-help}@
            goto retry_read; @\label{pread-internal-retry}@
        @\uline{\} else if (parent->status \& DirtyFlag) \{}@
            @\uline{PERSISTENT\_FLUSH(\&parent->status);}@
            @\uline{PERSISTENT\_FENCE();}@
            @\uline{parent->status = parent->status \& \textasciitilde{}DirtyFlag;}@
            @\uline{goto retry\_read;}@
        } else {@\label{pread-internal-line11}@
            return parent->status == SUCCESSFUL ?
                <val,val->new> : <val,val->old>;@\label{pread-internal-line12}@
}  }  }@\label{pr_i_end}@
\end{lstlisting}

\subsection{Memory Management}
\label{sec:mem-mgt-app}
We describe the memory management in the context of the \mcas\ implementation presented in Section~\ref{sec:algos}.

We use two thread-local lists for reclaiming operation descriptors:
One list is for descriptors that have been finalized, but not detached yet (\texttt{finalizedDescList}), and another is for
descriptors that have been detached but to which readers might still hold references (\texttt{detachedDescList}).

In general, our memory management scheme is similar to an RCU (read-copy-update) implementation~\cite{McKenney17, MS98}.
We start with a simple blocking scheme, extending it into a non-blocking one.
Each thread maintains an epoch number, incremented by the thread 
upon the entry to and before the exit from the \ttread\ and \mcas\ functions
(see, e.g., Lines~\ref{read-epoch-start} and~\ref{read-epoch-end} in Listing~\ref{list:algo}).
In \texttt{retireForCleanup} function (cf.\ Line~\ref{retire-for-cleanup} in Listing~\ref{list:algo}), a thread adds the given descriptor 
to \texttt{finalizedDescList}.
Once the size of this list reaches a certain threshold, the thread invokes a function similar semantically to \texttt{synchronize\_rcu()}~\cite{McKenney17}.
That is, it runs through all thread epochs, and waits for every epoch with an odd value (indicating that a thread
is inside the \ttread\ or \mcas\ functions) to advance.
Once all epochs are traversed, all descriptors currently in the \texttt{detachedDescList} list can be 
reclaimed (returned to the operating system or put into a list of available descriptors for reuse).
At the same time, all descriptors currently in the \texttt{finalizedDescList} list can be moved
to the \texttt{detachedDescList} list, after replacing pointers to those descriptors in corresponding memory locations with their actual values.

To elaborate on this last step, given an \texttt{MCASDescriptor} descriptor \texttt{d} that is about to be moved 
from \texttt{finalizedDescList} to \texttt{detachedDescList},
a thread runs through all the \texttt{WordDescriptor}s stored in \texttt{d}.
For every such \texttt{WordDescriptor} \texttt{w}, the thread checks whether \texttt{w->address} is equal to \texttt{d} and if so,
writes \texttt{w->old} or \texttt{w->new} into \texttt{w->address} according to the status of \texttt{d}.
The check and the write are done atomically using \cas{}.

The scheme presented so far is blocking---if a thread does not advance its epoch number, any thread will be unable 
to complete the traversal of epochs.
To avoid this issue, each thread may store a local copy of all thread epochs it has seen during the last traversal.
On its next epoch traversal, it compares the current and the previously seen epochs for each thread $t$, and if those
two are different, it infers that $t$ has made progress.
If progress is detected for all threads, 
any descriptor that was placed into \texttt{finalizedDescList} (\texttt{detachedDescList}) before 
the previous epoch traversal can be detached (reclaimed, respectively).

Note that while this scheme is non-blocking, a failure of a thread might prevent reclamation of \emph{any} memory associated with descriptors.
This is a common issue with epoch-based reclamation schemes~\cite{DHK16}, which could be resolved either by 
enhancing the scheme (e.g., as in~\cite{Brown15}) or by switching to a different scheme, e.g., one based on hazard pointers~\cite{DHK16,Michael04}.


\subsection{Managing Persistent Memory}
\label{sec:persistent-memory}
In this section, we show how the memory management mechanism described in Section~\ref{sec:mem-mgmt} is extended to manage persistent memory.
Upon recovery from a crash, any pending \pmcas\ operation is applied using the same algorithm as presented in Listing~\ref{list:algo}.
Pending \pmcas\ operations can be found by scanning allocated descriptors (e.g., if descriptors are allocated from a pool, similar to David et al.~\cite{david18}).
Moreover, since we assume the recovery is done by a single thread, we can immediately 
detach and recycle any finalized descriptor (after writing back the actual values into corresponding memory locations).
Therefore, when considering persistent memory, 
the only change required to support correct recycling of descriptors (in addition to using a persistent memory allocator) is
flushing all writes while detaching descriptors and introducing a persistent fence 
right before reclaiming descriptors from \texttt{detachedDescList}.
The fence is required to avoid a situation where a detached descriptor is recycled and a crash happens while 
the descriptor is being initialized with new values.
In this case, and if a fence is not used, 
some memory locations may still point to the descriptor (since updates to those locations might have not been persisted before the crash), 
while the descriptor may already be updated with new content.
Note, though, that the flushes and the fence take place off the critical path, 
therefore their impact on the performance of \pmcas\ is expected to be negligible.

\subsection{Efficient Reads}
\label{sec:reads}
Once a memory location has been modified by an \mcas\ operation, even if by a failed one, it would
refer to an operation descriptor until that descriptor is detached.
Until that happens, the latency of a read operation from that memory location would be increased as it would have to access  
an operation descriptor to determine the value that needs to be returned by the read.
The memory management mechanism as described above, however, would detach the descriptor only
as a part of an \mcas\ operation.
This might cause degraded performance for read-dominated workloads in which \mcas\ operations are rare.

To this end, we propose the following optimization for eventual removal of references to an operation descriptor 
and storing the corresponding value directly in the memory location as part of the read operation.
If a read operation finds a pointer to a finalized operation descriptor, it will generate a pseudo-random number\footnote{Generating 
a local pseudo-random number is a relatively inexpensive operation that requires only a few processor cycles (see, for instance, the generator in ASCYLIB~\cite{ascylibcode}.)}.
With a small probability, it will run a simplified version of the memory reclamation scheme described above.
Specifically, it will scan epochs of all other threads, and then change the contents of the memory location it attempts to read
to the actual value (using \cas{}).
(To avoid deadlock between two threads scanning epoch numbers, a thread may indicate that it is in the middle of an epoch scan
so that any descriptor can be detached, but not recycled at that time.)

\subsection{Performance in read- and update-heavy workloads}\label{sec:extra-evaluation}

Figures~\ref{fig:dll-appendix} and \ref{fig:btree-appendix} show performance results for the 50\%, 98\%, and 100\% reads workloads in the doubly-linked list and B+-tree benchmarks, respectively. All other settings are the same as in Section~\ref{sec:eval}.
These more extreme workloads largely magnify the performance effects demonstrated by the workloads included in the paper (see Section~\ref{sec:eval}). Namely, as the write ratio increases, so does the gap by which AOPT outperforms PMwCAS in cases that involve contention between concurrent \mcas{} operations (i.e., small and moderate lists and trees). With 100\% reads, our algorithm performs on par with or slightly trails behind PMwCAS at every contention level. Since the workload involves no update operations (and thus no \mcas{}es), the lower complexity of our \mcas{} operations does not factor into the results, whereas the higher overhead of the extra-level of indirection in our read operations does factor in. 

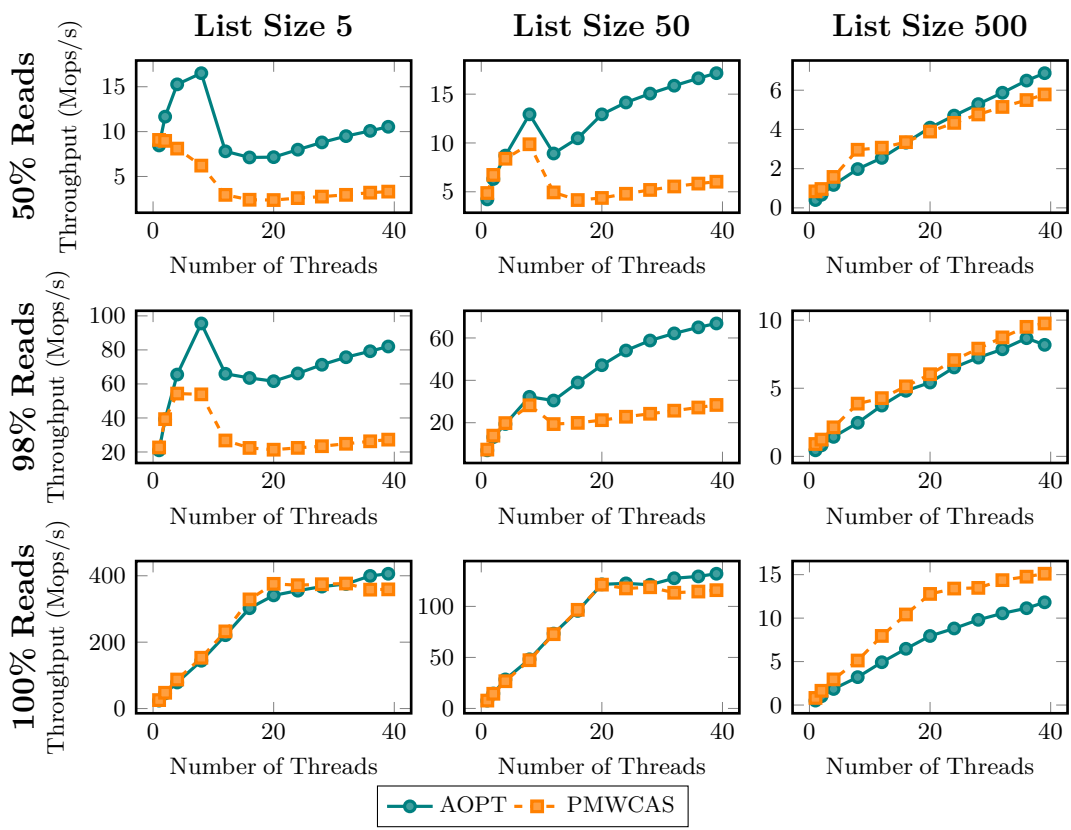
\begin{figure*}
\centering
\begin{tikzpicture}
	
    \begin{groupplot}[
        group style=    {group name=my plots,
                        group size=3 by 3,
                        horizontal sep=0.7cm,
                        vertical sep=1.3cm},
        height=3.6cm,
        width=5.2cm,
        ytick scale label code/.code={},
        legend columns=-1,
        xlabel={Number of Threads},
    ]
        
    \nextgroupplot[
        ylabel = {Throughput (Mops/s)},
        legend entries={AOPT,PMWCAS},
        legend to name=CommonLegend4,    
    ]
    \addplot table[x=Threads,y=AOPT] {data/dl-list/dl-list-list_size-5-search_pct-50-absolute-perf.log};
    \addplot table[x=Threads,y=PMWCAS] {data/dl-list/dl-list-list_size-5-search_pct-50-absolute-perf.log};
    \nextgroupplot
    \addplot table[x=Threads,y=AOPT] {data/dl-list/dl-list-list_size-50-search_pct-50-absolute-perf.log};
    \addplot table[x=Threads,y=PMWCAS] {data/dl-list/dl-list-list_size-50-search_pct-50-absolute-perf.log};
    \nextgroupplot
    \addplot table[x=Threads,y=AOPT] {data/dl-list/dl-list-list_size-500-search_pct-50-absolute-perf.log};
    \addplot table[x=Threads,y=PMWCAS] {data/dl-list/dl-list-list_size-500-search_pct-50-absolute-perf.log};
    
    \nextgroupplot[ylabel = {Throughput (Mops/s)}]
    \addplot table[x=Threads,y=AOPT] {data/dl-list/dl-list-list_size-5-search_pct-98-absolute-perf.log};
    \addplot table[x=Threads,y=PMWCAS] {data/dl-list/dl-list-list_size-5-search_pct-98-absolute-perf.log};
    \nextgroupplot
    \addplot table[x=Threads,y=AOPT] {data/dl-list/dl-list-list_size-50-search_pct-98-absolute-perf.log};
    \addplot table[x=Threads,y=PMWCAS] {data/dl-list/dl-list-list_size-50-search_pct-98-absolute-perf.log};
    \nextgroupplot
    \addplot table[x=Threads,y=AOPT] {data/dl-list/dl-list-list_size-500-search_pct-98-absolute-perf.log};
    \addplot table[x=Threads,y=PMWCAS] {data/dl-list/dl-list-list_size-500-search_pct-98-absolute-perf.log};  
    
    \nextgroupplot[ylabel = {Throughput (Mops/s)}]
    \addplot table[x=Threads,y=AOPT] {data/dl-list/dl-list-list_size-5-search_pct-100-absolute-perf.log};
    \addplot table[x=Threads,y=PMWCAS] {data/dl-list/dl-list-list_size-5-search_pct-100-absolute-perf.log};
    \nextgroupplot
    \addplot table[x=Threads,y=AOPT] {data/dl-list/dl-list-list_size-50-search_pct-100-absolute-perf.log};
    \addplot table[x=Threads,y=PMWCAS] {data/dl-list/dl-list-list_size-50-search_pct-100-absolute-perf.log};
    \nextgroupplot
    \addplot table[x=Threads,y=AOPT] {data/dl-list/dl-list-list_size-500-search_pct-100-absolute-perf.log};
    \addplot table[x=Threads,y=PMWCAS] {data/dl-list/dl-list-list_size-500-search_pct-100-absolute-perf.log};
    
    \end{groupplot}
	
	\node[anchor=south,rotate=90,yshift=1.2cm] at ($(my plots c1r1.west)$){{\Large\bf 50\% Reads}};
	\node[anchor=south,rotate=90,yshift=1.2cm] at ($(my plots c1r2.west)$){{\Large\bf 98\% Reads}};
	\node[anchor=south,rotate=90,yshift=1.2cm] at ($(my plots c1r3.west)$){{\Large\bf 100\% Reads}};

    \node[anchor=south,yshift=0.2cm] at ($(my plots c1r1.north)$){\Large \bf List Size 5};
    \node[anchor=south,yshift=0.2cm] at ($(my plots c2r1.north)$){\Large \bf List Size 50};
    \node[anchor=south,yshift=0.2cm] at ($(my plots c3r1.north)$){\Large \bf List Size 500};

\end{tikzpicture}
\\

\ref{CommonLegend4}
\caption{Doubly-linked list benchmark. Top row shows results for 50\% reads workload; middle row shows results for 98\% reads workload; bottom row shows results for 100\% reads workload. Each column corresponds to a different initial list size (5, 50 and 500 elements, respectively).}
\label{fig:dll-appendix}
\end{figure*}

\begin{figure*}
\centering
\begin{tikzpicture}
	
    \begin{groupplot}[
        group style=    {group name=my plots,
                        group size=3 by 3,
                        horizontal sep=0.7cm,
                        vertical sep=1.3cm},
        height=3.6cm,
        width=5.2cm,
        ytick scale label code/.code={},
        legend columns=-1,
        xlabel={Number of Threads},
    ]
        
    \nextgroupplot[
        ylabel = {Throughput (Mops/s)},
        legend entries={AOPT,PMWCAS},
        legend to name=CommonLegend5,
    ]
    \addplot table[x=Threads,y=AOPT] {data/bztree/bztree-tree_size-16-search_pct-50-absolute-perf.log};
    \addplot table[x=Threads,y=PMWCAS] {data/bztree/bztree-tree_size-16-search_pct-50-absolute-perf.log};
    \nextgroupplot
    \addplot table[x=Threads,y=AOPT] {data/bztree/bztree-tree_size-512-search_pct-50-absolute-perf.log};
    \addplot table[x=Threads,y=PMWCAS] {data/bztree/bztree-tree_size-512-search_pct-50-absolute-perf.log};
    \nextgroupplot
    \addplot table[x=Threads,y=AOPT] {data/bztree/bztree-tree_size-4000000-search_pct-50-absolute-perf.log};
    \addplot table[x=Threads,y=PMWCAS] {data/bztree/bztree-tree_size-4000000-search_pct-50-absolute-perf.log};
    
    \nextgroupplot[ylabel = {Throughput (Mops/s)}]
    \addplot table[x=Threads,y=AOPT] {data/bztree/bztree-tree_size-16-search_pct-98-absolute-perf.log};
    \addplot table[x=Threads,y=PMWCAS] {data/bztree/bztree-tree_size-16-search_pct-98-absolute-perf.log};
    \nextgroupplot
    \addplot table[x=Threads,y=AOPT] {data/bztree/bztree-tree_size-512-search_pct-98-absolute-perf.log};
    \addplot table[x=Threads,y=PMWCAS] {data/bztree/bztree-tree_size-512-search_pct-98-absolute-perf.log};
    \nextgroupplot
    \addplot table[x=Threads,y=AOPT] {data/bztree/bztree-tree_size-4000000-search_pct-98-absolute-perf.log};
    \addplot table[x=Threads,y=PMWCAS] {data/bztree/bztree-tree_size-4000000-search_pct-98-absolute-perf.log};    
    
    \nextgroupplot[ylabel = {Throughput (Mops/s)}]
    \addplot table[x=Threads,y=AOPT] {data/bztree/bztree-tree_size-16-search_pct-100-absolute-perf.log};
    \addplot table[x=Threads,y=PMWCAS] {data/bztree/bztree-tree_size-16-search_pct-100-absolute-perf.log};;
    \nextgroupplot
    \addplot table[x=Threads,y=AOPT] {data/bztree/bztree-tree_size-512-search_pct-100-absolute-perf.log};
    \addplot table[x=Threads,y=PMWCAS] {data/bztree/bztree-tree_size-512-search_pct-100-absolute-perf.log};
    \nextgroupplot
    \addplot table[x=Threads,y=AOPT] {data/bztree/bztree-tree_size-4000000-search_pct-100-absolute-perf.log};
    \addplot table[x=Threads,y=PMWCAS] {data/bztree/bztree-tree_size-4000000-search_pct-100-absolute-perf.log};
    
    \end{groupplot}
	
	\node[anchor=south,rotate=90,yshift=1cm] at ($(my plots c1r1.west)$){{\Large\bf 50\% Reads}};
	\node[anchor=south,rotate=90,yshift=1cm] at ($(my plots c1r2.west)$){{\Large\bf 98\% Reads}};
	\node[anchor=south,rotate=90,yshift=1cm] at ($(my plots c1r3.west)$){{\Large\bf 100\% Reads}};

    \node[anchor=south,yshift=0.2cm] at ($(my plots c1r1.north)$){\Large \bf Tree Size 16};
    \node[anchor=south,yshift=0.2cm] at ($(my plots c2r1.north)$){\Large \bf Tree Size 512};
    \node[anchor=south,yshift=0.2cm] at ($(my plots c3r1.north)$){\Large \bf Tree Size 4000000};

\end{tikzpicture}
\\

\ref{CommonLegend5}
\caption{B+-tree benchmark. Top row shows results for 50\% reads workload; middle row shows results for 98\% reads workload; bottom row shows results for 100\% reads workload. Each column corresponds to a different initial tree size (16, 512 and 4000000 elements, respectively).}
\label{fig:btree-appendix}
\end{figure*}
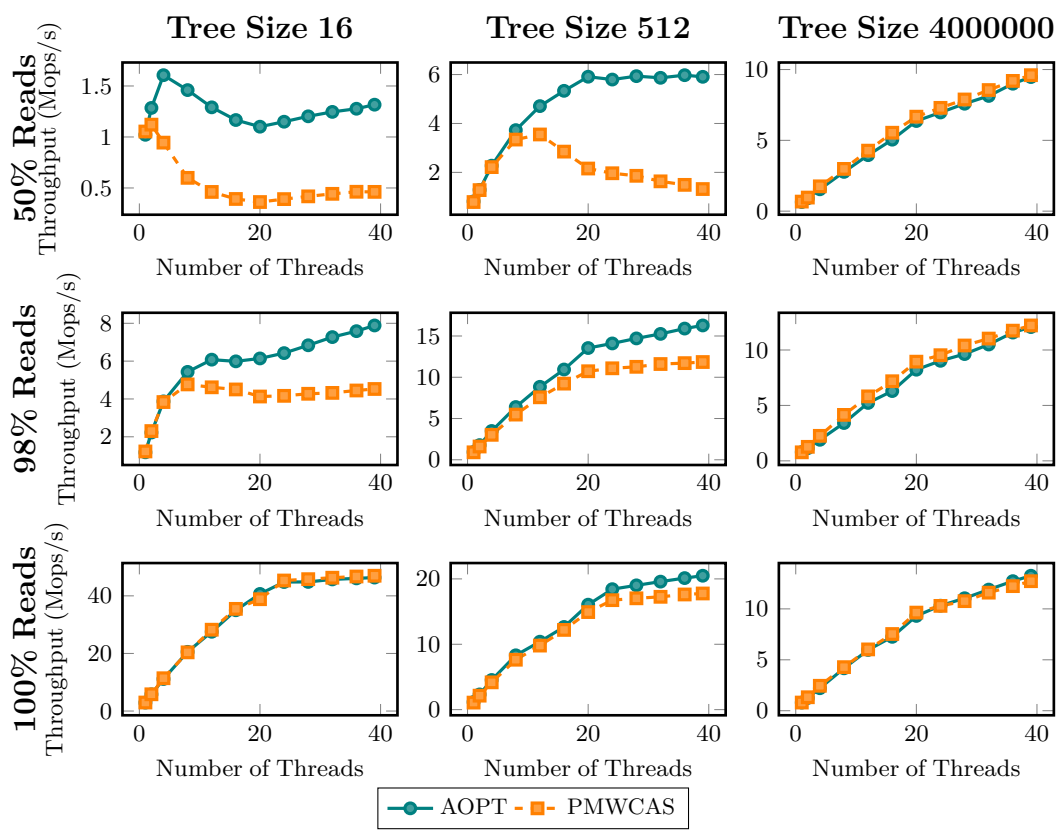

\subsection{Prior Work on Non-blocking \mcas{}}
\label{sec:related-non-blocking-mcas}

Israeli and Rappoport~\cite{israeli94} propose a lock-free and disjoint-access-parallel implementation based on \llsc{} and show how \llsc{} can be obtained from \cas{}. Their algorithm requires storing per-thread valid bits at each memory location, thus limiting the number of bits available for data. In the absence of contention, an a $k$-\cas{} requires $3k+2$ \cas{} instructions if using the \llsc{} implementation from \cas{} provided in the paper. In their implementation, uncontended reads (i.e., read operations that do not help concurrent \mcas{} operations) perform expensive atomic \textsf{LL} instructions, which can be emulated by writes to shared memory, thus limiting performance in common read-heavy workloads.

Anderson and Moir~\cite{anderson95} propose a wait-free implementation also based on \llsc{}. The strong progress guarantee comes with high space requirements: each memory word needs to be followed contiguously by an  auxiliary word containing information needed to help complete an ongoing operation on the memory word.

Moir~\cite{moir97} simplifies~\cite{anderson95} considerably by removing the requirement of wait-freedom. Instead, his algorithm is conditionally wait-free: it is lock-free and provides a means to communicate with an external helping mechanism which may cancel \mcas{} operations that are no longer required to complete.

Harris et al.~\cite{harris02} introduce a lock-free algorithm based on \cas{} operations. In order to avoid the ABA problem, the algorithm uses a double-compare-single-swap primitive (implemented using two \cas{} instructions, in the absence of contention) to make each target word point to a global \mcas{} descriptor while ensuring that the descriptor is still active. In order to distinguish between values and descriptors, the two least-significant bits are reserved in each word. In total, a $k$-word \mcas{} uses $3k+1$ \cas{} instructions in the uncontended case. 

Ha and Tsigas~\cite{ha03,ha04} provide lock-free algorithms which measure the amount of contention on \mcas{} target words and react by dynamically choosing the best helping policy.

Attiya and Hillel~\cite{attiya11} give a lock-free implementation using \cas{} and D\cas{} that requires $6k+2$ \cas{} instructions for a $k$-word \mcas{} in the uncontended case. To avoid the ABA problem, this algorithm stores a tag with each pointer which it atomically increments every time the pointer changes. The algorithm uses a conflict-resolution scheme in which contending operations decide whether to help or reset one another based on how many locations each operation acquired before the conflict was detected (preference is given to operations that own more locations). Their implementation does not provide a separate {\em read} operation.

Sundell~\cite{sundell11} proposes a scheme that uses $2k+1$ \cas{} instructions for a $k$-word \mcas{} (in the absence of contention). An \mcas{} operation first uses \cas{} to acquire ownership of each target word, changes the status using a \cas{} and then uses \cas{} to write the final values back into the target word. The algorithm is wait-free under the assumption that there is a bound on the number of \mcas{} operations with equal old and new values. 

Feldman et al.~\cite{feldman15} propose an algorithm that is both wait-free and ABA-free. In their helping mechanism, a thread actively announces if it is blocked (i.e., if it fails to complete due to concurrent \mcas{} operations), relying on contending operations to help it to complete.

\subparagraph{Restricted and extended multi-word operations.}

Previous work has explored other operations that atomically read and modify multiple words. These operations are either more general or more restricted than \mcas{}.

Luchangco, Moir and Shavit~\cite{luchangco03} present an obstruction-free
implementation of a ``$k$-compare-single-swap'', which compares on $k$ words but only modifies one word (more restricted than \mcas{}). Their algorithm is based on \llsc{}, for which they give an obstruction-free implementation from \cas{}.

Brown et al.~\cite{brown13} introduce extensions to \llsc{} called \textsc{LLX/SCX}, which are more general than $k$-compare-single-swap, but more restricted than \mcas{}.
\textsc{LLX/SCX} primitives operate on sets of data records, each comprising several words. 
\textsc{SCX} allows modifying a single word of a data record, conditional on the fact that no data record in a specified set was modified since \textsc{LLX} was last performed on it. 
Furthermore, \textsc{SCX} allows finalizing a subset of the data records, preventing them from being modified again.
While \textsc{LLX/SCX} and \mcas{} can be used to solve similar problems, \mcas{} is more generic, as it allows modifying $k$ words atomically, whereas \textsc{LLX/SCX} only allow modifying a single word.

Timnat et al.~\cite{timnat15} propose an extension of \mcas{} called \textsc{MCMS} (Multiple Compare Multiple Swap), which also allows addresses to be compared without being swapped (more general than \mcas{}). They provide implementations of \textsc{MCMS} based on HTM and on the algorithm by Harris et al.~\cite{harris02}.
\end{arxiv}

\end{document}